\documentclass[%
 aip,amsmath,amssymb,
preprint,%
]{revtex4-1}
\usepackage{verbatim}
\usepackage{soul}
\usepackage{graphicx}
\usepackage{dcolumn}
\usepackage{bm}
\usepackage{xspace}
\usepackage{xcolor}
\newcommand{\am}{Auger-Meitner\xspace}
\newcommand{\pa}{phonon-assisted\xspace}
\newcommand{\amu}{cm\(^6\)s\(^{-1}\)\xspace}
\newcommand{\oee}{\(10^{18}\) cm\(^{-3}\)\xspace}

\newcommand{\beginsupplement}{
            \setcounter{table}{0}
            \renewcommand{\thetable}{S\arabic{table}}
            \setcounter{figure}{0}
            \renewcommand{\thefigure}{S\arabic{figure}}
}

\begin{document}

\title{Strain Effects on Auger-Meitner Recombination in Silicon}
\author{Kyle Bushick}
\affiliation{Department of Materials Science and Engineering, University of Michigan, Ann Arbor, MI, USA}
\affiliation{Materials Science Division, Lawrence Livermore National Laboratory, Livermore, CA, USA}
\author{Emmanouil Kioupakis}
 \email{kioup@umich.edu}
\affiliation{Department of Materials Science and Engineering, University of Michigan, Ann Arbor, MI, USA}
\date{13 December 2023}

\begin{abstract}
We study the effects of compressive and tensile biaxial strain on direct and \pa Auger-Meitner recombination (AMR) in silicon using first-principles calculations. We find that the application of strain has a non-trivial effect on the AMR rate. For most AMR processes, the application of strain increases the AMR rate. However, the recombination rate for the AMR process involving two holes and one electron is suppressed by 38\% under tensile strain. We further analyze the specific phonon contributions that mediate the \pa AMR mechanism, demonstrating the increased anisotropy under strain. Our results indicate that the application of tensile strain increases the lifetime of minority electron carriers in \textit{p}-type silicon, and can be leveraged to improve the efficiency of silicon devices. 
\end{abstract}

\maketitle

\am recombination (AMR) is a non-radiative recombination process intrinsic to semiconductor materials that involves three carriers. In AMR, an electron-hole pair recombines across the band gap, transferring the excess energy through the Coulomb interaction to a third carrier, either an electron or a hole, which is itself promoted to a higher energy state. If the third carrier is an electron, we denote this as the electron-electron-hole (\(eeh\)) process, while if the third carrier is a hole, we denote this as the hole-hole-electron (\(hhe\)) process. Furthermore, AMR can occur either in a direct manner, where energy and momentum are strictly conserved between the participating carriers, or in an indirect (\pa) manner, where one of the carriers can absorb or emit a phonon, relaxing the momentum conservation constraints and enabling excitations to a broader range of high-energy states. Historically, this process has been known as Auger recombination, but the renaming to \am aims to recognize the contributions of Lise Meitner in the discovery of the effect.\cite{Matsakis2019} In silicon, AMR acts as a loss mechanism, limiting the efficiency of numerous devices including solar cells,\cite{Green1984,Tiedje1984,Kerr2003,Su2023} transistors,\cite{Shibib1979,Tyagi1983} and diodes.\cite{Shibib1979,Leilaeioun2016} 

In many functional applications, applying a small amount of strain is a technique that can be used to tune the material properties and improve device performance.\cite{Tsutsui2019,Dai2019,Miao2022} This is the case both for silicon and for other materials.\cite{Omi2003,Ieong2004,Paul2004,Chidambaram2006,Rodl2015,Meesala2018,Cai2021} However, there is little work in the literature on the effects of strain on AMR in silicon, a deficiency we aim to address in our work. Ultimately, it is of interest to modulate the AMR rate in devices {\textendash} either to mitigate its effects as a non-radiative loss mechanism\cite{Itsuno2012,Bae2013,Piprek2016} or to intentionally increase the generation of hot carriers.\cite{Singh2019,Livache2022,Du2022} In the case of silicon devices, we are mostly interested in the former. Considering AMR is an intrinsic material property that is independent of defects present in the material, engineering solutions that aim to improve the chemical purity or structural quality are not viable methods to control the AMR rate. However, affecting the underlying electronic structure is a viable option for controlling the AMR rate. As has been shown in previous work on InGaAs/InGaAsP quantum well lasers, the application of compressive strain is a viable strategy to suppress AMR by affecting the valence band curvature and therefore the availability of holes that satisfy both momentum and energy conservation constraints.\cite{Lui1994} While this specific AMR-suppression mechanism does not translate to bulk silicon {\textendash} for which direct \(hhe\) AMR is negligible compared to other AMR pathways\cite{Bushick2023} {\textendash} there may be other modifications to the band structure that affect the AMR rate. Specifically, the application of biaxial strain allows one to alter the valley and band degeneracy of electrons and holes. In the case of the electrons in silicon, the six-fold valley degeneracy is split, with the valleys along the direction of crystal compression (four in-plane for compressive and two out-of-plane for tensile biaxial strain) lowering in energy. For holes, the valence band maximum is at the \(\Gamma\)-point, so there is no valley degeneracy, however the application of strain reduces the band degeneracy with one and two band(s) lowering in energy under the application of compressive and tensile biaxial strain, respectively. These effects are illustrated in Fig. \ref{fig:bands} (we cover the strain effects on the phonon dispersion in the Supplementary Material). In the context of AMR, these alterations to the band structure lead to non-trivial impacts on the recombination rates, emphasizing the need for first-principles characterization techniques. 

\begin{figure*}[ht]
\includegraphics{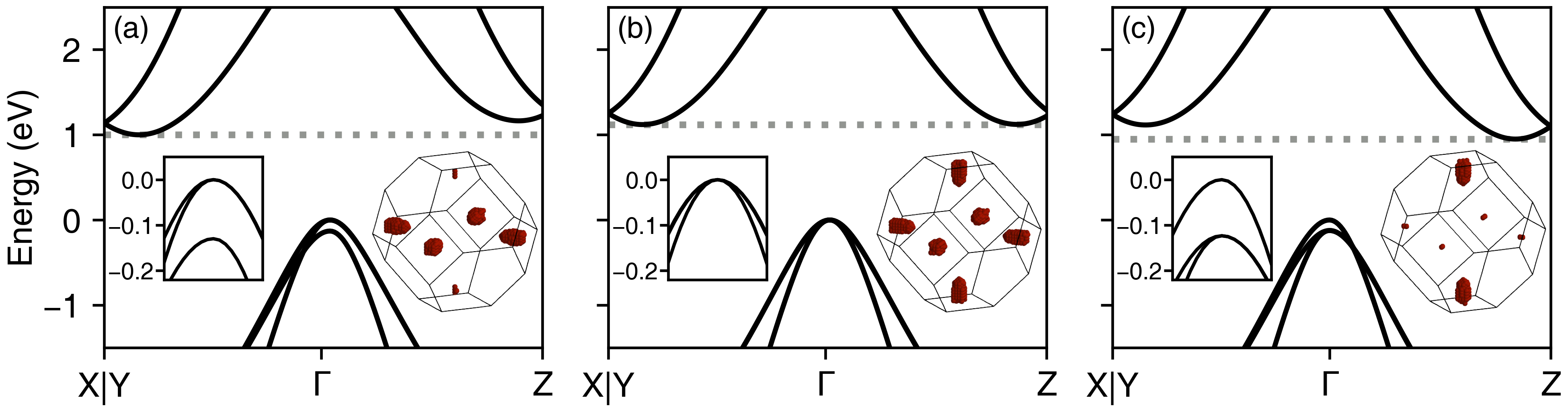}
\caption{Effects of biaxial strain on the band structure of silicon. 1\% compressive (a), unstrained (b), and 1\% tensile (c) conditions are shown. The left inset shows a zoomed view of the valence band edges, while the right inset shows the relative occupation of the conduction band valleys in the first Brillouin zone. The grey dotted lines are guides to the eye to show the conduction band energy splitting.}
\label{fig:bands}
\end{figure*}

In this Letter, we apply our first-principles methodology to characterize the effects of compressive and tensile biaxial strain on both the direct and \pa AMR rate in crystalline silicon. We show that the AMR coefficient increases for both \(eeh\) and \(hhe\) AMR under compressive strain, as well as \(eeh\) AMR under tensile strain. Notably, we find a \(\sim\)40\% reduction in the \(hhe\) AMR coefficient under tensile strain. We also analyze the contributions of different phonon modes to the \pa AMR process under various strain conditions. Finally, we discuss how our findings can inform the application of strain to improve silicon device performance. 

Our calculation methodology uses eigenvalues, wave functions, and electron-phonon coupling matrix elements generated by density functional theory (DFT) and density functional perturbation theory (DFPT) calculations, which we obtain using the \texttt{Quantum ESPRESSO} code\cite{Giannozzi2009,Giannozzi2017,Giannozzi2020} within the local density approximation (LDA) for the exchange-correlation functional.\cite{Ceperley1980,Perdew1981} We use a Troullier-Martins\cite{Troullier1991} norm-conserving pseudo-potential throughout our calculations. We use a relaxed lattice parameter of \(a=5.379\) {\AA} for the unstrained structure. To obtain the biaxially strained structures, we adjust the in-plane lattice constants (X and Y)  by \(\mp1\%\) and allow the out-of-plane Z direction to relax. This gives us an out-of-plane lattice parameter of 5.420 {\AA} for the compressive biaxial strain condition and 5.337 {\AA} for the tensile biaxial strain condition. 

We chose to study this magnitude of strain because it is readily achievable in real systems and leads to sizable valley (\(>170\) meV) and band (\(>125\) meV) energy splitting for the electrons and holes, respectively (Fig. \ref{fig:bands}a,c). It is important to achieve splitting much larger than kT (\(\sim\)26 meV) to ensure that carriers selectively occupy the lower-energy states, since otherwise the AMR rates would be approximately equal to those of unstrained Si. While we expect a smooth transition from the unstrained to the 1\% strain AMR values as energy splitting goes from zero to \(>\)kT, once a substantial polarization of the state occupancies is achieved, further increases to the strain are not expected to yield large variations in the AMR coefficients. Since the spin-orbit splitting is only \(\sim\)45 meV in silicon,\cite{Bona1985,Ponce2021}  the band-structure splitting effects are dominated by strain. Furthermore, work by Zacharias, Scheffler, and Carbogno on unstrained silicon has shown that the electron-phonon renormalization of the band gap is approximately 60 meV at 0 K and less than 100 meV at 300 K.\cite{Zacharias2020} The explicit inclusion of these effects is beyond the scope of our work, but we expect the electron-phonon renormalization to be similar for the energy-split states, and thus it should only have a minor effect on the AMR coefficients in strained silicon. The 1\% strain value is therefore sufficiently high enough to achieve the full effects of strain on the AMR coefficients, while simultaneously being low enough to realize experimentally.

We calculated the LDA eigenvalues on an \(8\times8\times8\) Brillouin-zone sampling grid and interpolated them to fine grids using the maximally localized Wannier function method\cite{Marzari2012} and the \texttt{Wannier90} code.\cite{Pizzi2020} While we performed G\textsubscript{0}W\textsubscript{0} (GW) calculations to obtain quasiparticle corrections to the LDA eigenenergies for all three structures, we found that in all cases the GW corrections were effectively rigid shifts to the band gaps (Fig. S2). To avoid the added complexity of performing Wannier-function interpolation on the GW quasiparticle energies to arbitrary \textbf{k}-points, we opted to rigidly shift the band gaps to 1.00 eV for the compressed system, 1.12 eV for the unstrained system, and 0.95 eV for the tensile system. These values were chosen because they are consistent with the experimental value for unstrained silicon,\cite{Bludau1974} the LDA band gap differences between the strained and unstrained structures, as well as with experimental band-gap values for strained silicon reported in the literature.\cite{Munguia2008} To calculate the AMR coefficients, we used the same converged Brillouin zone sampling grids of \(50\times50\times50\) and \(\delta\)-function broadenings of 0.1-0.2 eV discussed in our previous work.\cite{Bushick2023} We use a temperature of 300 K and free-carrier concentration of \oee throughout our calculations. Under these conditions,  Coulomb-enhancement effects to the AMR rate that arise from many-body interactions are weak and can be safely ignored.\cite{Richter2012,Bushick2023}

\begin{figure}[ht]
\includegraphics{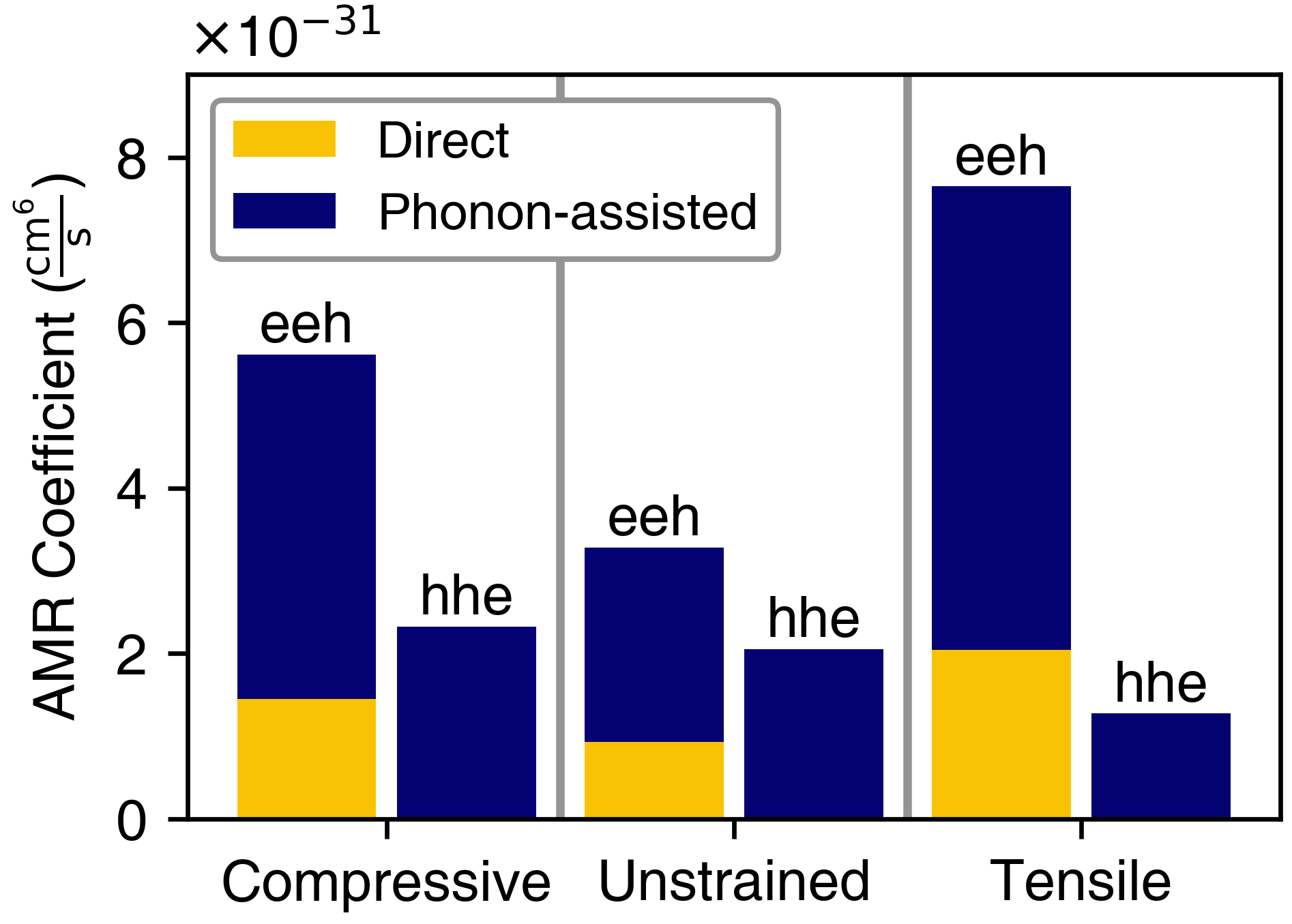}
\caption{Effects of strain on the \(eeh\) and \(hhe\) AMR coefficients in silicon. The contributions from the direct process are shown in gold, while the contributions from the \pa process are shown in blue. Direct \(hhe\) is  forbidden by momentum conservation under all strain conditions, and only the  \(hhe\) process is reduced by the application of tensile strain.}
\label{fig:amr}
\end{figure}

To perform our AMR calculations for strained silicon, we applied first- and second-order time dependent perturbation theory in the Fermi's golden rule framework, as outlined in our previous work on unstrained silicon.\cite{Bushick2023} From these calculations we obtain the AMR rate, \(R=\frac{dN}{dt}=CVn^3\), where the carrier density \(n = \frac{N}{V}\) denotes the number of free carriers \(N\) per volume \(V\), and \(C\) is the AMR coefficient. For direct AMR, we make no assumptions about the Brillouin-zone distribution of the free carriers as a function of strain conditions, and there are therefore no differences in our approach compared to those discussed in Refs. \citenum{Bushick2023} and \citenum{Kioupakis2015}. As we show in the insets of Fig. \ref{fig:bands} and Fig. S3 in greater detail, the majority (\(>\)97\%) of electrons occupy the lower-energy conduction band valley (while the holes are primarily restricted to the top valence band). Given this information and the assumptions needed for the computational tractability of the \pa calculations (that the periodic part of the wave functions of carriers are approximately equal to those of the nearest band extremum), we restrict \(N^\mathbb{C}_{valley}=4\) for compressive strain and \(N^\mathbb{C}_{valley}=2\) for tensile strain. Similarly, we restrict \(N^\mathbb{V}_{band}=2\) for compressive strain and \(N^\mathbb{V}_{band}=1\) for tensile strain. This captures the effects of strain on the band structure while keeping the calculation computationally feasible. We do not include the effects of spin-orbit splitting in our calculations, as the energy splitting due to strain is approximately \(3\times\) stronger, as we discuss above.

\begin{table}[]
\centering
\begin{tabular}{ccccc}
Strain Condition & \(C_{eeh,dir}\) & \(C_{eeh,pa}\) & \(C_{hhe,dir}\) & \(C_{hhe,pa}\) \\ \hline\hline
\(-1\%\) strain  & 1.45            & 4.17           & 0.0000293       & 2.32           \\
No strain        & 0.932           & 2.35           & 0.0000194       & 2.05           \\
\(+1\%\) strain  & 2.04            & 5.62           & 0.00000489      & 1.27          
\end{tabular}
\caption{Calculated AMR coefficients under different strain conditions. All values are in units of \(\times10^{-31}\) \amu.}
\label{tab:amrc}
\end{table}

Using our computational approach outlined above, we report the \(eeh\) and \(hhe\) AMR rates across the three strain conditions (compressive, unstrained, and tensile) in Fig. \ref{fig:amr} and Table \ref{tab:amrc}. We break down the total AMR coefficient in terms of the direct and \pa contributions. As expected, our results for the unstrained system are consistent with past work, both theoretical and experimental.\cite{Dziewior1977,Hacker1994,Govoni2011,Bushick2023} The similarity between our unstrained values and past results indicates the validity of using the LDA eigenvalues with a rigid shift in place of the GW quasiparticle energies. Under the application of compressive strain, we find \(C_{eeh,dir}\) increases by \(56\%\), \(C_{eeh,pa}\) increases by \(77\%\), \(C_{hhe,dir}\) increases by \(51\%\), and \(C_{hhe,pa}\) increases by \(13\%\). In the case of tensile strain, we find \(C_{eeh,dir}\) increases by \(119\%\), \(C_{eeh,pa}\) increases by \(139\%\), \(C_{hhe,dir}\) decreases by \(75\%\), and \(C_{hhe,pa}\) decreases by \(38\%\). Under all strain conditions, the direct \(hhe\) AMR process is effectively forbidden due to momentum conservation, and so \(hhe\) AMR is entirely accounted for by the \pa process. As we show, applying strain generally increases the AMR rate, with the exception of \(hhe\) under tensile strain. This is largely consistent with previous work on strained germanium, which found that both direct and phonon-assisted AMR increased under strain, though these results coincide with an indirect-to-direct band gap transition in germanium, which we attribute as the cause for the differing behavior.\cite{Dominici2016}

\begin{figure}[ht]
\includegraphics[width=6.5in]{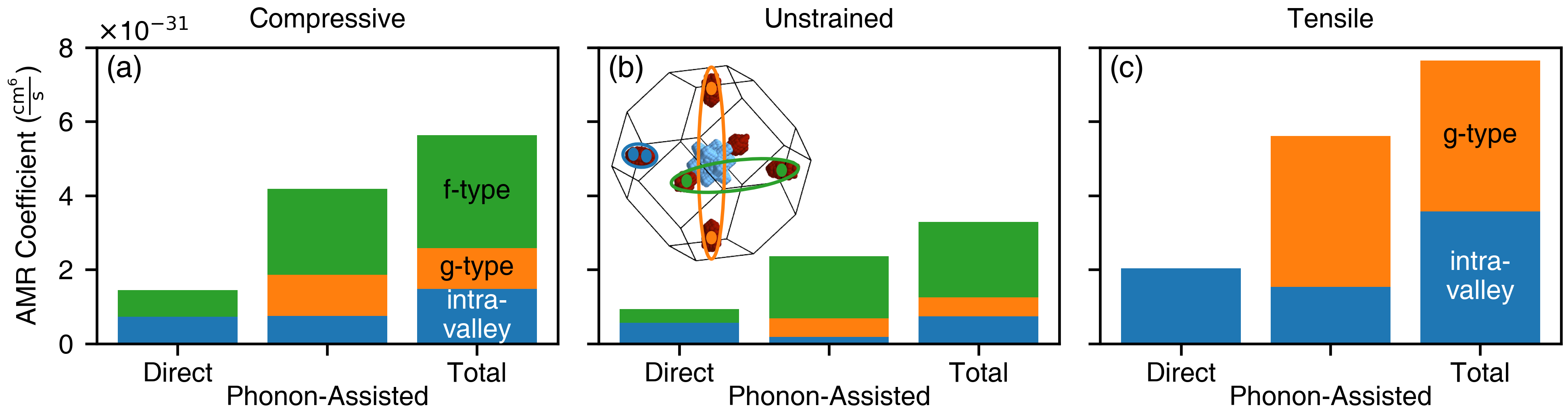}
\caption{Breakdown of \(eeh\) AMR for direct and \pa processes in strained silicon. The broken valley degeneracy affects the effective carrier concentration and available scattering pathways for the low-energy electrons. The most notable effect is the elimination of \textit{f}-type AMR under tensile strain. The inset plot in (b) serves as a visual key for the different recombination types, with the colors matching the bar colors (intravalley = blue, \textit{g}-type = orange, \textit{f}-type = green).}
\label{fig:valleys}
\end{figure}

Two competing effects alter the AMR rate due to the symmetry and degeneracy breaking induced by strain: (1) selective occupation of bands with reweighted matrix elements, and (2) a reduced number of recombination channels. The interplay between these two effects is clearly seen in Fig. \ref{fig:valleys}, where \textit{f}-type (\(eeh\)) AMR is eliminated in the tensile condition, but the strength of the \textit{g}-type and intravalley processes is significantly enhanced. Note that since both holes for \(hhe\) AMR occupy the same \(\Gamma\) valley, there is no distinction between their initial valley arrangement. Further discussion of the valley analysis is included in the Supplementary Material. Since we expect a decrease in AMR due to the reduction of recombination channels but consistently observe an increase in the AMR rates, we conclude that the selective band occupancy dominates the effects of strain on AMR. However, the non-trivial effects on the AMR coefficient that result from the competition and interplay between these strain-induced mechanisms highlights the need for the first-principles approach employed in this investigation.

\begin{figure}[ht]
\includegraphics{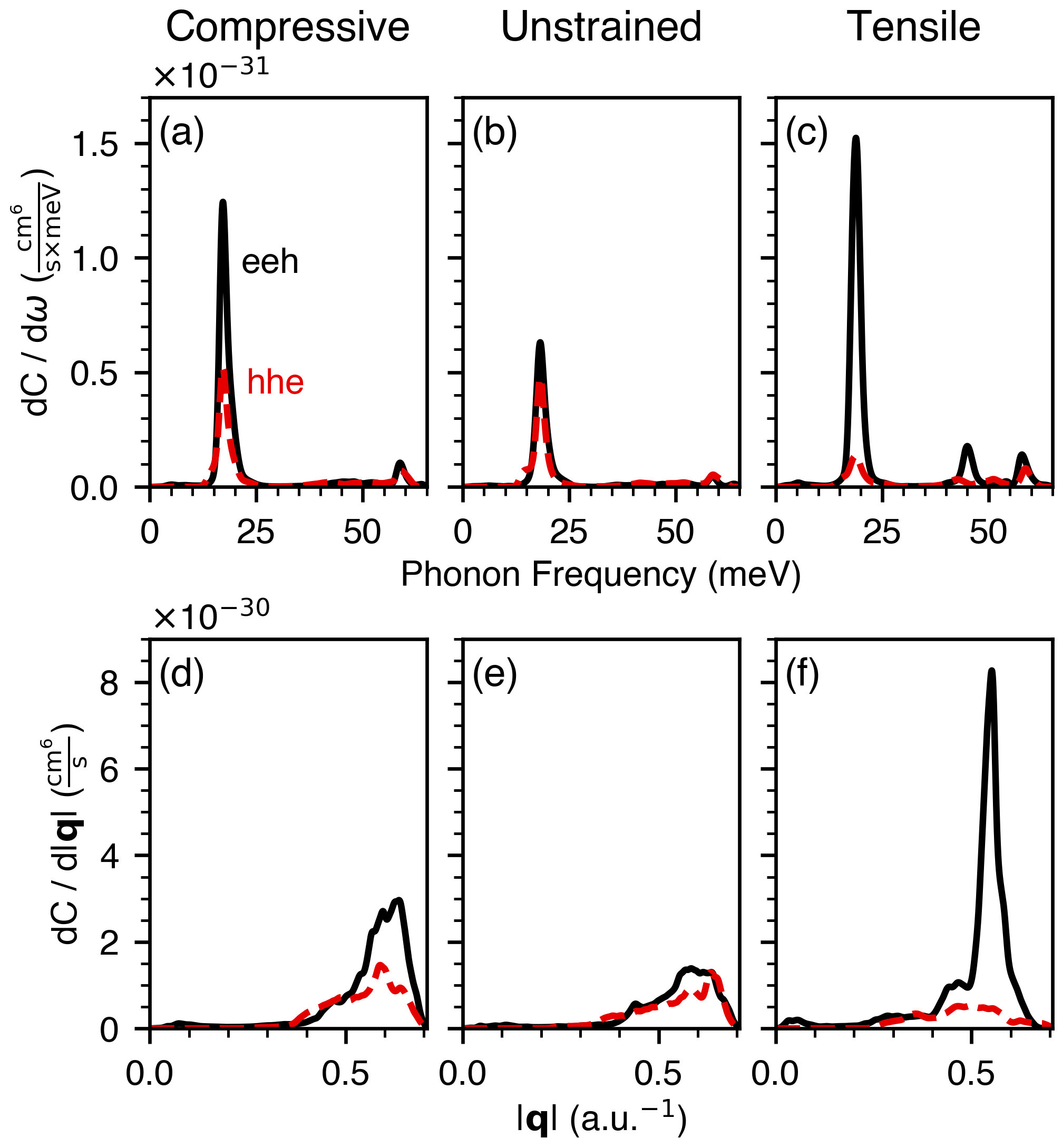}
\caption{Phonon decomposition of the \pa \(eeh\) (solid black) and \(hhe\) (dashed red) AMR coefficients. (a-c) show the dependence of the AMR coefficient on the phonon frequency for the compressive, unstrained, and tensile systems. The distribution is mostly consistent across the strain conditions, with the \(hhe\) process becoming relatively weaker compared to the \(eeh\) process under strain. The  \(eeh\) process under tensile strain shows the emergence of a new contribution by phonons at 45 meV. (d-f) show the AMR distribution as a function of phonon wave-vector magnitude across the three strain conditions. The qualitative observations are consistent with the energy decomposition, though the tensile \(eeh\) process also shows a prominent peak corresponding to \(\mathbf{q}\approx\Delta_Z\).}
\label{fig:ph_analysis}
\end{figure}

An additional benefit of leveraging first-principles calculations is the ability to obtain insights into the underlying mechanisms of AMR. Figure \ref{fig:ph_analysis} details the contributions of different phonons to the overall \pa AMR process, expressed in terms of distributions over phonon energy and phonon momentum. The integration of these distributions yields the total AMR coefficient. The three-dimensional momentum distribution is shown in Fig. S4. We find that the distribution over phonon modes are reweighted through the application of strain. These effects are most clearly seen in the \(eeh\) process under tensile strain, but are present under all conditions. Of particular note for the phonon-assisted \(eeh\) AMR under tensile strain is the emergence of a third peak at a phonon energy of 45 meV, as well as a prominent renormalization at the phonon wave vector magnitude of 0.55 a.u.\textsuperscript{-1}, which corresponds to phonons with \(\mathbf{q}\approx\Delta_Z\), i.e., the momentum of the low-energy out-of-plane electron valley. The magnitude of this phonon mode contribution also indicates a larger matrix element for this transition under tensile strain. The effects of compressive strain on the distribution are less pronounced, which is intuitive given that the band structure more closely resembles that of the unstrained system and no recombination pathways are forbidden, unlike in the tensile case. We also analyze the excited electron distribution in the Supplementary Material (Fig. S5), highlighting the anisotropy of the direct \(eeh\) AMR process. 

Our results not only provide insights into the microscopic mechanisms governing AMR in strained silicon, but also into the potential application of strain to improve silicon device performance. Indeed, silicon devices are incredibly varied and technologically critical. They encompass diverse electronic and optoelectronic applications from solar cells to transistors, playing a critical role in both our energy and computing infrastructure. In many of these applications, AMR is known to act as a loss mechanism, which reduces their overall efficiency and performance. This is notable in solar cells, transistors, thyristors, and \textit{p-n} junctions more generally.\cite{Shibib1979,Tyagi1983,Tiedje1984,Green1984,Kerr2003,Sze2006,Su2023} Unfortunately, it is challenging to adjust one intrinsic material property (e.g., band gap, mobility, carrier lifetime) by strain in isolation. While we focus on the effects of strain on AMR in this Letter, we note the fact that both compressive and tensile biaxial strain decreases the band gap of silicon. Other works have also discussed the effects of strain on the carrier mobility, which is a property of great importance to many devices.\cite{Yu2008} Specifically, applying strain on the order of \(\mp1\%\) leads to a reduction of the band gap on the order of 150 meV and non-trivial changes to the electron and hole mobilities. While the application of strain mostly leads to an increase in the AMR rate, we have shown that 1\% tensile strain reduces the \(hhe\) AMR rate by 38\%. Considering the combined effects of tensile strain on increasing the in-plane electron mobility\cite{Yu2008} and suppressing the \(hhe\) AMR rate (i.e., increasing the lifetime of electrons in \textit{p}-type materials), strain engineering may be a route to increase the performance of devices utilizing minority electron carriers in a \textit{p}-type region. We note, however, that tensile strain also reduces the out-of-plane electron mobility.\cite{Yu2008} Unfortunately, it is unlikely this approach could be directly applied to the \textit{p}-type regions of silicon solar cells, where thicknesses on the order of 200 \(\mu\)m preclude the coherent application of strain. The performance of other devices such as transistors, however, which are significantly smaller, may be able to benefit by the application of tensile strain. Indeed, any application where the ideality factor ranges between 2/3 and 1 (i.e., AMR is the lifetime-limiting process) could leverage this strain engineering approach to improve efficiency.\cite{Green1984,Leilaeioun2016} Future studies may also investigate the non-trivial dependence of the AMR coefficients on the magnitude of strain, optimizing the interplay between the various effects for specific applications.

In summary, we apply our first-principles methodology to investigate the effects of compressive and tensile biaxial strain on \am recombination in silicon. The application of strain alters the electronic structure and leads to an increase in the AMR coefficient in all cases except \(hhe\) AMR under tensile biaxial strain, where AMR is reduced by \(\sim\)40\%. We demonstrate that the effects of strain are non-trivial and give rise to qualitatively distinct behavior for the \pa AMR process. Our findings inform the application of strain to affect device performance, highlighted by increased minority carrier lifetimes in \textit{p}-type regions. 

See the Supplementary Material for additional information on (1) the strain effects on the phonon dispersion (2) the quasiparticle corrections and band gap shifts, (3) the carrier distribution changes under strain, (4) valley analysis of the electrons in the \(eeh\) process, (5) three-dimensional plots of the phonon contributions, and (6) distribution of excited carriers. 

The authors have no conflicts to disclose.

The data that support the findings of this study are available from the corresponding author upon reasonable request. 

\begin{acknowledgments}
We thank David Young for useful discussions. This work is supported as part of the Computational Materials Sciences Program funded by the U.S. Department of Energy, Office of Science, Basic Energy Sciences under Award No. DE-SC0020129. This work used resources of the National Energy Research Scientific Computing (NERSC) Center, a DOE Office of Science User Facility supported under Contract No. DE-AC02–05CH11231. K.B. acknowledges the support of the U.S. Department of Energy, Office of Science, Office of Advanced Scientific Computing Research, Department of Energy Computational Science Graduate Fellowship under Award Number DE-SC0020347. K.B.'s editing of this manuscript was performed under the auspices of the U.S. Department of Energy by Lawrence Livermore National Laboratory under Contract DE-AC52-07NA27344. 

\end{acknowledgments}
\bibliography{main.bib}

\begin{thebibliography}{47}%
\makeatletter
\providecommand \@ifxundefined [1]{%
 \@ifx{#1\undefined}
}%
\providecommand \@ifnum [1]{%
 \ifnum #1\expandafter \@firstoftwo
 \else \expandafter \@secondoftwo
 \fi
}%
\providecommand \@ifx [1]{%
 \ifx #1\expandafter \@firstoftwo
 \else \expandafter \@secondoftwo
 \fi
}%
\providecommand \natexlab [1]{#1}%
\providecommand \enquote  [1]{``#1''}%
\providecommand \bibnamefont  [1]{#1}%
\providecommand \bibfnamefont [1]{#1}%
\providecommand \citenamefont [1]{#1}%
\providecommand \href@noop [0]{\@secondoftwo}%
\providecommand \href [0]{\begingroup \@sanitize@url \@href}%
\providecommand \@href[1]{\@@startlink{#1}\@@href}%
\providecommand \@@href[1]{\endgroup#1\@@endlink}%
\providecommand \@sanitize@url [0]{\catcode `\\12\catcode `\$12\catcode `\&12\catcode `\#12\catcode `\^12\catcode `\_12\catcode `\%12\relax}%
\providecommand \@@startlink[1]{}%
\providecommand \@@endlink[0]{}%
\providecommand \url  [0]{\begingroup\@sanitize@url \@url }%
\providecommand \@url [1]{\endgroup\@href {#1}{\urlprefix }}%
\providecommand \urlprefix  [0]{URL }%
\providecommand \Eprint [0]{\href }%
\providecommand \doibase [0]{http://dx.doi.org/}%
\providecommand \selectlanguage [0]{\@gobble}%
\providecommand \bibinfo  [0]{\@secondoftwo}%
\providecommand \bibfield  [0]{\@secondoftwo}%
\providecommand \translation [1]{[#1]}%
\providecommand \BibitemOpen [0]{}%
\providecommand \bibitemStop [0]{}%
\providecommand \bibitemNoStop [0]{.\EOS\space}%
\providecommand \EOS [0]{\spacefactor3000\relax}%
\providecommand \BibitemShut  [1]{\csname bibitem#1\endcsname}%
\let\auto@bib@innerbib\@empty
\bibitem [{\citenamefont {Matsakis}\ \emph {et~al.}(2019)\citenamefont {Matsakis}, \citenamefont {Coster}, \citenamefont {Laster},\ and\ \citenamefont {Sime}}]{Matsakis2019}%
  \BibitemOpen
  \bibfield  {author} {\bibinfo {author} {\bibfnamefont {D.}~\bibnamefont {Matsakis}}, \bibinfo {author} {\bibfnamefont {A.}~\bibnamefont {Coster}}, \bibinfo {author} {\bibfnamefont {B.}~\bibnamefont {Laster}}, \ and\ \bibinfo {author} {\bibfnamefont {R.}~\bibnamefont {Sime}},\ }\bibfield  {title} {\enquote {\bibinfo {title} {A renaming proposal: “{T}he {A}uger–{M}eitner effect”},}\ }\href {\doibase 10.1063/PT.3.4281} {\bibfield  {journal} {\bibinfo  {journal} {Physics Today}\ }\textbf {\bibinfo {volume} {72}},\ \bibinfo {pages} {10--11} (\bibinfo {year} {2019})}\BibitemShut {NoStop}%
\bibitem [{\citenamefont {Green}(1984)}]{Green1984}%
  \BibitemOpen
  \bibfield  {author} {\bibinfo {author} {\bibfnamefont {M.}~\bibnamefont {Green}},\ }\bibfield  {title} {\enquote {\bibinfo {title} {Limits on the open-circuit voltage and efficiency of silicon solar cells imposed by intrinsic {A}uger processes},}\ }\href {\doibase 10.1109/T-ED.1984.21588} {\bibfield  {journal} {\bibinfo  {journal} {IEEE Transactions on Electron Devices}\ }\textbf {\bibinfo {volume} {31}},\ \bibinfo {pages} {671--678} (\bibinfo {year} {1984})}\BibitemShut {NoStop}%
\bibitem [{\citenamefont {Tiedje}\ \emph {et~al.}(1984)\citenamefont {Tiedje}, \citenamefont {Yablonovitch}, \citenamefont {Cody},\ and\ \citenamefont {Brooks}}]{Tiedje1984}%
  \BibitemOpen
  \bibfield  {author} {\bibinfo {author} {\bibfnamefont {T.}~\bibnamefont {Tiedje}}, \bibinfo {author} {\bibfnamefont {E.}~\bibnamefont {Yablonovitch}}, \bibinfo {author} {\bibfnamefont {G.}~\bibnamefont {Cody}}, \ and\ \bibinfo {author} {\bibfnamefont {B.}~\bibnamefont {Brooks}},\ }\bibfield  {title} {\enquote {\bibinfo {title} {Limiting efficiency of silicon solar cells},}\ }\href {\doibase 10.1109/T-ED.1984.21594} {\bibfield  {journal} {\bibinfo  {journal} {IEEE Transactions on Electron Devices}\ }\textbf {\bibinfo {volume} {31}},\ \bibinfo {pages} {711--716} (\bibinfo {year} {1984})}\BibitemShut {NoStop}%
\bibitem [{\citenamefont {Kerr}, \citenamefont {Cuevas},\ and\ \citenamefont {Campbell}(2003)}]{Kerr2003}%
  \BibitemOpen
  \bibfield  {author} {\bibinfo {author} {\bibfnamefont {M.~J.}\ \bibnamefont {Kerr}}, \bibinfo {author} {\bibfnamefont {A.}~\bibnamefont {Cuevas}}, \ and\ \bibinfo {author} {\bibfnamefont {P.}~\bibnamefont {Campbell}},\ }\bibfield  {title} {\enquote {\bibinfo {title} {Limiting efficiency of crystalline silicon solar cells due to {C}oulomb-enhanced {A}uger recombination},}\ }\href {\doibase 10.1002/pip.464} {\bibfield  {journal} {\bibinfo  {journal} {Progress in Photovoltaics: Research and Applications}\ }\textbf {\bibinfo {volume} {11}},\ \bibinfo {pages} {97--104} (\bibinfo {year} {2003})}\BibitemShut {NoStop}%
\bibitem [{\citenamefont {Su}\ \emph {et~al.}(2023)\citenamefont {Su}, \citenamefont {Lin}, \citenamefont {Wang}, \citenamefont {Tang}, \citenamefont {Xue}, \citenamefont {Li}, \citenamefont {Xu},\ and\ \citenamefont {Gao}}]{Su2023}%
  \BibitemOpen
  \bibfield  {author} {\bibinfo {author} {\bibfnamefont {Q.}~\bibnamefont {Su}}, \bibinfo {author} {\bibfnamefont {H.}~\bibnamefont {Lin}}, \bibinfo {author} {\bibfnamefont {G.}~\bibnamefont {Wang}}, \bibinfo {author} {\bibfnamefont {H.}~\bibnamefont {Tang}}, \bibinfo {author} {\bibfnamefont {C.}~\bibnamefont {Xue}}, \bibinfo {author} {\bibfnamefont {Z.}~\bibnamefont {Li}}, \bibinfo {author} {\bibfnamefont {X.}~\bibnamefont {Xu}}, \ and\ \bibinfo {author} {\bibfnamefont {P.}~\bibnamefont {Gao}},\ }\bibfield  {title} {\enquote {\bibinfo {title} {Limiting-{E}fficiency {A}ssessment on {A}dvanced {C}rystalline {S}ilicon {S}olar {C}ells with {A}uger {I}deality {F}actor and {W}afer {T}hickness {M}odifications},}\ }\href {\doibase 10.22541/au.167930468.87567881/v1} {\bibfield  {journal} {\bibinfo  {journal} {Authorea}\ } (\bibinfo {year} {2023}),\ 10.22541/au.167930468.87567881/v1}\BibitemShut {NoStop}%
\bibitem [{\citenamefont {Shibib}, \citenamefont {Lindholm},\ and\ \citenamefont {Fossum}(1979)}]{Shibib1979}%
  \BibitemOpen
  \bibfield  {author} {\bibinfo {author} {\bibfnamefont {M.}~\bibnamefont {Shibib}}, \bibinfo {author} {\bibfnamefont {F.}~\bibnamefont {Lindholm}}, \ and\ \bibinfo {author} {\bibfnamefont {J.}~\bibnamefont {Fossum}},\ }\bibfield  {title} {\enquote {\bibinfo {title} {Auger recombination in heavily doped shallow-emitter silicon p-n-junction solar cells, diodes, and transistors},}\ }\href {\doibase 10.1109/T-ED.1979.19555} {\bibfield  {journal} {\bibinfo  {journal} {IEEE Transactions on Electron Devices}\ }\textbf {\bibinfo {volume} {26}},\ \bibinfo {pages} {1104--1106} (\bibinfo {year} {1979})}\BibitemShut {NoStop}%
\bibitem [{\citenamefont {Tyagi}\ and\ \citenamefont {Overstraeten}(1983)}]{Tyagi1983}%
  \BibitemOpen
  \bibfield  {author} {\bibinfo {author} {\bibfnamefont {M.}~\bibnamefont {Tyagi}}\ and\ \bibinfo {author} {\bibfnamefont {R.~V.}\ \bibnamefont {Overstraeten}},\ }\bibfield  {title} {\enquote {\bibinfo {title} {Minority carrier recombination in heavily-doped silicon},}\ }\href {\doibase 10.1016/0038-1101(83)90174-0} {\bibfield  {journal} {\bibinfo  {journal} {Solid-State Electronics}\ }\textbf {\bibinfo {volume} {26}},\ \bibinfo {pages} {577--597} (\bibinfo {year} {1983})}\BibitemShut {NoStop}%
\bibitem [{\citenamefont {Leilaeioun}\ and\ \citenamefont {Holman}(2016)}]{Leilaeioun2016}%
  \BibitemOpen
  \bibfield  {author} {\bibinfo {author} {\bibfnamefont {M.}~\bibnamefont {Leilaeioun}}\ and\ \bibinfo {author} {\bibfnamefont {Z.~C.}\ \bibnamefont {Holman}},\ }\bibfield  {title} {\enquote {\bibinfo {title} {Accuracy of expressions for the fill factor of a solar cell in terms of open-circuit voltage and ideality factor},}\ }\href {\doibase 10.1063/1.4962511} {\bibfield  {journal} {\bibinfo  {journal} {Journal of Applied Physics}\ }\textbf {\bibinfo {volume} {120}},\ \bibinfo {pages} {123111} (\bibinfo {year} {2016})}\BibitemShut {NoStop}%
\bibitem [{\citenamefont {Tsutsui}\ \emph {et~al.}(2019)\citenamefont {Tsutsui}, \citenamefont {Mochizuki}, \citenamefont {Loubet}, \citenamefont {Bedell},\ and\ \citenamefont {Sadana}}]{Tsutsui2019}%
  \BibitemOpen
  \bibfield  {author} {\bibinfo {author} {\bibfnamefont {G.}~\bibnamefont {Tsutsui}}, \bibinfo {author} {\bibfnamefont {S.}~\bibnamefont {Mochizuki}}, \bibinfo {author} {\bibfnamefont {N.}~\bibnamefont {Loubet}}, \bibinfo {author} {\bibfnamefont {S.~W.}\ \bibnamefont {Bedell}}, \ and\ \bibinfo {author} {\bibfnamefont {D.~K.}\ \bibnamefont {Sadana}},\ }\bibfield  {title} {\enquote {\bibinfo {title} {Strain engineering in functional materials},}\ }\href {\doibase 10.1063/1.5075637} {\bibfield  {journal} {\bibinfo  {journal} {AIP Advances}\ }\textbf {\bibinfo {volume} {9}},\ \bibinfo {pages} {030701} (\bibinfo {year} {2019})}\BibitemShut {NoStop}%
\bibitem [{\citenamefont {Dai}, \citenamefont {Liu},\ and\ \citenamefont {Zhang}(2019)}]{Dai2019}%
  \BibitemOpen
  \bibfield  {author} {\bibinfo {author} {\bibfnamefont {Z.}~\bibnamefont {Dai}}, \bibinfo {author} {\bibfnamefont {L.}~\bibnamefont {Liu}}, \ and\ \bibinfo {author} {\bibfnamefont {Z.}~\bibnamefont {Zhang}},\ }\bibfield  {title} {\enquote {\bibinfo {title} {Strain {E}ngineering of 2{D} {M}aterials: {I}ssues and {O}pportunities at the {I}nterface},}\ }\href {\doibase 10.1002/adma.201805417} {\bibfield  {journal} {\bibinfo  {journal} {Advanced Materials}\ }\textbf {\bibinfo {volume} {31}},\ \bibinfo {pages} {1805417} (\bibinfo {year} {2019})}\BibitemShut {NoStop}%
\bibitem [{\citenamefont {Miao}\ \emph {et~al.}(2022)\citenamefont {Miao}, \citenamefont {Zhao}, \citenamefont {Zhang}, \citenamefont {Shi},\ and\ \citenamefont {Zhang}}]{Miao2022}%
  \BibitemOpen
  \bibfield  {author} {\bibinfo {author} {\bibfnamefont {Y.}~\bibnamefont {Miao}}, \bibinfo {author} {\bibfnamefont {Y.}~\bibnamefont {Zhao}}, \bibinfo {author} {\bibfnamefont {S.}~\bibnamefont {Zhang}}, \bibinfo {author} {\bibfnamefont {R.}~\bibnamefont {Shi}}, \ and\ \bibinfo {author} {\bibfnamefont {T.}~\bibnamefont {Zhang}},\ }\bibfield  {title} {\enquote {\bibinfo {title} {Strain {E}ngineering: {A} {B}oosting {S}trategy for {P}hotocatalysis},}\ }\href {\doibase 10.1002/adma.202200868} {\bibfield  {journal} {\bibinfo  {journal} {Advanced Materials}\ }\textbf {\bibinfo {volume} {34}},\ \bibinfo {pages} {2200868} (\bibinfo {year} {2022})}\BibitemShut {NoStop}%
\bibitem [{\citenamefont {Omi}\ \emph {et~al.}(2003)\citenamefont {Omi}, \citenamefont {Bottomley}, \citenamefont {Homma},\ and\ \citenamefont {Ogino}}]{Omi2003}%
  \BibitemOpen
  \bibfield  {author} {\bibinfo {author} {\bibfnamefont {H.}~\bibnamefont {Omi}}, \bibinfo {author} {\bibfnamefont {D.~J.}\ \bibnamefont {Bottomley}}, \bibinfo {author} {\bibfnamefont {Y.}~\bibnamefont {Homma}}, \ and\ \bibinfo {author} {\bibfnamefont {T.}~\bibnamefont {Ogino}},\ }\bibfield  {title} {\enquote {\bibinfo {title} {Wafer-scale strain engineering on silicon for fabrication of ultimately controlled nanostructures},}\ }\href {\doibase 10.1103/PhysRevB.67.115302} {\bibfield  {journal} {\bibinfo  {journal} {Physical Review B}\ }\textbf {\bibinfo {volume} {67}},\ \bibinfo {pages} {115302} (\bibinfo {year} {2003})}\BibitemShut {NoStop}%
\bibitem [{\citenamefont {Ieong}\ \emph {et~al.}(2004)\citenamefont {Ieong}, \citenamefont {Doris}, \citenamefont {Kedzierski}, \citenamefont {Rim},\ and\ \citenamefont {Yang}}]{Ieong2004}%
  \BibitemOpen
  \bibfield  {author} {\bibinfo {author} {\bibfnamefont {M.}~\bibnamefont {Ieong}}, \bibinfo {author} {\bibfnamefont {B.}~\bibnamefont {Doris}}, \bibinfo {author} {\bibfnamefont {J.}~\bibnamefont {Kedzierski}}, \bibinfo {author} {\bibfnamefont {K.}~\bibnamefont {Rim}}, \ and\ \bibinfo {author} {\bibfnamefont {M.}~\bibnamefont {Yang}},\ }\bibfield  {title} {\enquote {\bibinfo {title} {Silicon {D}evice {S}caling to the {S}ub-10-nm {R}egime},}\ }\href {\doibase 10.1126/science.1100731} {\bibfield  {journal} {\bibinfo  {journal} {Science}\ }\textbf {\bibinfo {volume} {306}},\ \bibinfo {pages} {2057--2060} (\bibinfo {year} {2004})}\BibitemShut {NoStop}%
\bibitem [{\citenamefont {Paul}(2004)}]{Paul2004}%
  \BibitemOpen
  \bibfield  {author} {\bibinfo {author} {\bibfnamefont {D.~J.}\ \bibnamefont {Paul}},\ }\bibfield  {title} {\enquote {\bibinfo {title} {Si/{S}i{G}e heterostructures: from material and physics to devices and circuits},}\ }\href {\doibase 10.1088/0268-1242/19/10/R02} {\bibfield  {journal} {\bibinfo  {journal} {Semiconductor Science and Technology}\ }\textbf {\bibinfo {volume} {19}},\ \bibinfo {pages} {R75--R108} (\bibinfo {year} {2004})}\BibitemShut {NoStop}%
\bibitem [{\citenamefont {Chidambaram}\ \emph {et~al.}(2006)\citenamefont {Chidambaram}, \citenamefont {Bowen}, \citenamefont {Chakravarthi}, \citenamefont {Machala},\ and\ \citenamefont {Wise}}]{Chidambaram2006}%
  \BibitemOpen
  \bibfield  {author} {\bibinfo {author} {\bibfnamefont {P.}~\bibnamefont {Chidambaram}}, \bibinfo {author} {\bibfnamefont {C.}~\bibnamefont {Bowen}}, \bibinfo {author} {\bibfnamefont {S.}~\bibnamefont {Chakravarthi}}, \bibinfo {author} {\bibfnamefont {C.}~\bibnamefont {Machala}}, \ and\ \bibinfo {author} {\bibfnamefont {R.}~\bibnamefont {Wise}},\ }\bibfield  {title} {\enquote {\bibinfo {title} {Fundamentals of silicon material properties for successful exploitation of strain engineering in modern {CMOS} manufacturing},}\ }\href {\doibase 10.1109/TED.2006.872912} {\bibfield  {journal} {\bibinfo  {journal} {IEEE Transactions on Electron Devices}\ }\textbf {\bibinfo {volume} {53}},\ \bibinfo {pages} {944--964} (\bibinfo {year} {2006})}\BibitemShut {NoStop}%
\bibitem [{\citenamefont {Rödl}\ \emph {et~al.}(2015)\citenamefont {Rödl}, \citenamefont {Sander}, \citenamefont {Bechstedt}, \citenamefont {Vidal}, \citenamefont {Olsson}, \citenamefont {Laribi},\ and\ \citenamefont {Guillemoles}}]{Rodl2015}%
  \BibitemOpen
  \bibfield  {author} {\bibinfo {author} {\bibfnamefont {C.}~\bibnamefont {Rödl}}, \bibinfo {author} {\bibfnamefont {T.}~\bibnamefont {Sander}}, \bibinfo {author} {\bibfnamefont {F.}~\bibnamefont {Bechstedt}}, \bibinfo {author} {\bibfnamefont {J.}~\bibnamefont {Vidal}}, \bibinfo {author} {\bibfnamefont {P.}~\bibnamefont {Olsson}}, \bibinfo {author} {\bibfnamefont {S.}~\bibnamefont {Laribi}}, \ and\ \bibinfo {author} {\bibfnamefont {J.-F.}\ \bibnamefont {Guillemoles}},\ }\bibfield  {title} {\enquote {\bibinfo {title} {Wurtzite silicon as a potential absorber in photovoltaics: {T}ailoring the optical absorption by applying strain},}\ }\href {\doibase 10.1103/PhysRevB.92.045207} {\bibfield  {journal} {\bibinfo  {journal} {Physical Review B}\ }\textbf {\bibinfo {volume} {92}},\ \bibinfo {pages} {045207} (\bibinfo {year} {2015})}\BibitemShut {NoStop}%
\bibitem [{\citenamefont {Meesala}\ \emph {et~al.}(2018)\citenamefont {Meesala}, \citenamefont {Sohn}, \citenamefont {Pingault}, \citenamefont {Shao}, \citenamefont {Atikian}, \citenamefont {Holzgrafe}, \citenamefont {Gündoğan}, \citenamefont {Stavrakas}, \citenamefont {Sipahigil}, \citenamefont {Chia}, \citenamefont {Evans}, \citenamefont {Burek}, \citenamefont {Zhang}, \citenamefont {Wu}, \citenamefont {Pacheco}, \citenamefont {Abraham}, \citenamefont {Bielejec}, \citenamefont {Lukin}, \citenamefont {Atatüre},\ and\ \citenamefont {Lončar}}]{Meesala2018}%
  \BibitemOpen
  \bibfield  {author} {\bibinfo {author} {\bibfnamefont {S.}~\bibnamefont {Meesala}}, \bibinfo {author} {\bibfnamefont {Y.-I.}\ \bibnamefont {Sohn}}, \bibinfo {author} {\bibfnamefont {B.}~\bibnamefont {Pingault}}, \bibinfo {author} {\bibfnamefont {L.}~\bibnamefont {Shao}}, \bibinfo {author} {\bibfnamefont {H.~A.}\ \bibnamefont {Atikian}}, \bibinfo {author} {\bibfnamefont {J.}~\bibnamefont {Holzgrafe}}, \bibinfo {author} {\bibfnamefont {M.}~\bibnamefont {Gündoğan}}, \bibinfo {author} {\bibfnamefont {C.}~\bibnamefont {Stavrakas}}, \bibinfo {author} {\bibfnamefont {A.}~\bibnamefont {Sipahigil}}, \bibinfo {author} {\bibfnamefont {C.}~\bibnamefont {Chia}}, \bibinfo {author} {\bibfnamefont {R.}~\bibnamefont {Evans}}, \bibinfo {author} {\bibfnamefont {M.~J.}\ \bibnamefont {Burek}}, \bibinfo {author} {\bibfnamefont {M.}~\bibnamefont {Zhang}}, \bibinfo {author} {\bibfnamefont {L.}~\bibnamefont {Wu}}, \bibinfo {author} {\bibfnamefont {J.~L.}\ \bibnamefont {Pacheco}}, \bibinfo {author} {\bibfnamefont {J.}~\bibnamefont
  {Abraham}}, \bibinfo {author} {\bibfnamefont {E.}~\bibnamefont {Bielejec}}, \bibinfo {author} {\bibfnamefont {M.~D.}\ \bibnamefont {Lukin}}, \bibinfo {author} {\bibfnamefont {M.}~\bibnamefont {Atatüre}}, \ and\ \bibinfo {author} {\bibfnamefont {M.}~\bibnamefont {Lončar}},\ }\bibfield  {title} {\enquote {\bibinfo {title} {Strain engineering of the silicon-vacancy center in diamond},}\ }\href {\doibase 10.1103/PhysRevB.97.205444} {\bibfield  {journal} {\bibinfo  {journal} {Physical Review B}\ }\textbf {\bibinfo {volume} {97}},\ \bibinfo {pages} {205444} (\bibinfo {year} {2018})}\BibitemShut {NoStop}%
\bibitem [{\citenamefont {Cai}\ \emph {et~al.}(2021)\citenamefont {Cai}, \citenamefont {Wang}, \citenamefont {He}, \citenamefont {Liu}, \citenamefont {Xiong}, \citenamefont {Liu},\ and\ \citenamefont {Zhang}}]{Cai2021}%
  \BibitemOpen
  \bibfield  {author} {\bibinfo {author} {\bibfnamefont {W.}~\bibnamefont {Cai}}, \bibinfo {author} {\bibfnamefont {J.}~\bibnamefont {Wang}}, \bibinfo {author} {\bibfnamefont {Y.}~\bibnamefont {He}}, \bibinfo {author} {\bibfnamefont {S.}~\bibnamefont {Liu}}, \bibinfo {author} {\bibfnamefont {Q.}~\bibnamefont {Xiong}}, \bibinfo {author} {\bibfnamefont {Z.}~\bibnamefont {Liu}}, \ and\ \bibinfo {author} {\bibfnamefont {Q.}~\bibnamefont {Zhang}},\ }\bibfield  {title} {\enquote {\bibinfo {title} {Strain-{M}odulated {P}hotoelectric {R}esponses from a {F}lexible \(\alpha\)-{I}n\textsubscript{2}{S}e\textsubscript{3}/3{R} {M}o{S}\textsubscript{2} {H}eterojunction},}\ }\href {\doibase 10.1007/s40820-020-00584-1} {\bibfield  {journal} {\bibinfo  {journal} {Nano-Micro Letters}\ }\textbf {\bibinfo {volume} {13}},\ \bibinfo {pages} {74} (\bibinfo {year} {2021})}\BibitemShut {NoStop}%
\bibitem [{\citenamefont {Itsuno}, \citenamefont {Phillips},\ and\ \citenamefont {Velicu}(2012)}]{Itsuno2012}%
  \BibitemOpen
  \bibfield  {author} {\bibinfo {author} {\bibfnamefont {A.~M.}\ \bibnamefont {Itsuno}}, \bibinfo {author} {\bibfnamefont {J.~D.}\ \bibnamefont {Phillips}}, \ and\ \bibinfo {author} {\bibfnamefont {S.}~\bibnamefont {Velicu}},\ }\bibfield  {title} {\enquote {\bibinfo {title} {Design of an {A}uger-{S}uppressed {U}nipolar {H}g{C}d{T}e {NB}\(\nu\){N} {P}hotodetector},}\ }\href {\doibase 10.1007/s11664-012-1992-y} {\bibfield  {journal} {\bibinfo  {journal} {Journal of Electronic Materials}\ }\textbf {\bibinfo {volume} {41}},\ \bibinfo {pages} {2886--2892} (\bibinfo {year} {2012})}\BibitemShut {NoStop}%
\bibitem [{\citenamefont {Bae}\ \emph {et~al.}(2013)\citenamefont {Bae}, \citenamefont {Padilha}, \citenamefont {Park}, \citenamefont {McDaniel}, \citenamefont {Robel}, \citenamefont {Pietryga},\ and\ \citenamefont {Klimov}}]{Bae2013}%
  \BibitemOpen
  \bibfield  {author} {\bibinfo {author} {\bibfnamefont {W.~K.}\ \bibnamefont {Bae}}, \bibinfo {author} {\bibfnamefont {L.~A.}\ \bibnamefont {Padilha}}, \bibinfo {author} {\bibfnamefont {Y.-S.}\ \bibnamefont {Park}}, \bibinfo {author} {\bibfnamefont {H.}~\bibnamefont {McDaniel}}, \bibinfo {author} {\bibfnamefont {I.}~\bibnamefont {Robel}}, \bibinfo {author} {\bibfnamefont {J.~M.}\ \bibnamefont {Pietryga}}, \ and\ \bibinfo {author} {\bibfnamefont {V.~I.}\ \bibnamefont {Klimov}},\ }\bibfield  {title} {\enquote {\bibinfo {title} {Controlled {A}lloying of the {C}ore–{S}hell {I}nterface in {C}d{S}e/{C}d{S} {Q}uantum {D}ots for {S}uppression of {A}uger {R}ecombination},}\ }\href {\doibase 10.1021/nn4002825} {\bibfield  {journal} {\bibinfo  {journal} {ACS Nano}\ }\textbf {\bibinfo {volume} {7}},\ \bibinfo {pages} {3411--3419} (\bibinfo {year} {2013})}\BibitemShut {NoStop}%
\bibitem [{\citenamefont {Piprek}(2016)}]{Piprek2016}%
  \BibitemOpen
  \bibfield  {author} {\bibinfo {author} {\bibfnamefont {J.}~\bibnamefont {Piprek}},\ }\bibfield  {title} {\enquote {\bibinfo {title} {Analysis of efficiency limitations in high-power {InGaN/GaN} laser diodes},}\ }\href {\doibase 10.1007/s11082-016-0727-3} {\bibfield  {journal} {\bibinfo  {journal} {Optical and Quantum Electronics}\ }\textbf {\bibinfo {volume} {48}},\ \bibinfo {pages} {471} (\bibinfo {year} {2016})}\BibitemShut {NoStop}%
\bibitem [{\citenamefont {Singh}\ \emph {et~al.}(2019)\citenamefont {Singh}, \citenamefont {Liu}, \citenamefont {Lim}, \citenamefont {Robel},\ and\ \citenamefont {Klimov}}]{Singh2019}%
  \BibitemOpen
  \bibfield  {author} {\bibinfo {author} {\bibfnamefont {R.}~\bibnamefont {Singh}}, \bibinfo {author} {\bibfnamefont {W.}~\bibnamefont {Liu}}, \bibinfo {author} {\bibfnamefont {J.}~\bibnamefont {Lim}}, \bibinfo {author} {\bibfnamefont {I.}~\bibnamefont {Robel}}, \ and\ \bibinfo {author} {\bibfnamefont {V.~I.}\ \bibnamefont {Klimov}},\ }\bibfield  {title} {\enquote {\bibinfo {title} {Hot-electron dynamics in quantum dots manipulated by spin-exchange {A}uger interactions},}\ }\href {\doibase 10.1038/s41565-019-0548-1} {\bibfield  {journal} {\bibinfo  {journal} {Nature Nanotechnology}\ }\textbf {\bibinfo {volume} {14}},\ \bibinfo {pages} {1035--1041} (\bibinfo {year} {2019})}\BibitemShut {NoStop}%
\bibitem [{\citenamefont {Livache}\ \emph {et~al.}(2022)\citenamefont {Livache}, \citenamefont {Kim}, \citenamefont {Jin}, \citenamefont {Kozlov}, \citenamefont {Fedin},\ and\ \citenamefont {Klimov}}]{Livache2022}%
  \BibitemOpen
  \bibfield  {author} {\bibinfo {author} {\bibfnamefont {C.}~\bibnamefont {Livache}}, \bibinfo {author} {\bibfnamefont {W.~D.}\ \bibnamefont {Kim}}, \bibinfo {author} {\bibfnamefont {H.}~\bibnamefont {Jin}}, \bibinfo {author} {\bibfnamefont {O.~V.}\ \bibnamefont {Kozlov}}, \bibinfo {author} {\bibfnamefont {I.}~\bibnamefont {Fedin}}, \ and\ \bibinfo {author} {\bibfnamefont {V.~I.}\ \bibnamefont {Klimov}},\ }\bibfield  {title} {\enquote {\bibinfo {title} {High-efficiency photoemission from magnetically doped quantum dots driven by multi-step spin-exchange {A}uger ionization},}\ }\href {\doibase 10.1038/s41566-022-00989-x} {\bibfield  {journal} {\bibinfo  {journal} {Nature Photonics}\ }\textbf {\bibinfo {volume} {16}},\ \bibinfo {pages} {433--440} (\bibinfo {year} {2022})}\BibitemShut {NoStop}%
\bibitem [{\citenamefont {Du}\ \emph {et~al.}(2022)\citenamefont {Du}, \citenamefont {Yin}, \citenamefont {Xie}, \citenamefont {Sun}, \citenamefont {Fang}, \citenamefont {Wang}, \citenamefont {Li}, \citenamefont {Xiao}, \citenamefont {Yang}, \citenamefont {Zhang}, \citenamefont {Wang}, \citenamefont {Chen}, \citenamefont {Yin},\ and\ \citenamefont {Zheng}}]{Du2022}%
  \BibitemOpen
  \bibfield  {author} {\bibinfo {author} {\bibfnamefont {S.}~\bibnamefont {Du}}, \bibinfo {author} {\bibfnamefont {J.}~\bibnamefont {Yin}}, \bibinfo {author} {\bibfnamefont {H.}~\bibnamefont {Xie}}, \bibinfo {author} {\bibfnamefont {Y.}~\bibnamefont {Sun}}, \bibinfo {author} {\bibfnamefont {T.}~\bibnamefont {Fang}}, \bibinfo {author} {\bibfnamefont {Y.}~\bibnamefont {Wang}}, \bibinfo {author} {\bibfnamefont {J.}~\bibnamefont {Li}}, \bibinfo {author} {\bibfnamefont {D.}~\bibnamefont {Xiao}}, \bibinfo {author} {\bibfnamefont {X.}~\bibnamefont {Yang}}, \bibinfo {author} {\bibfnamefont {S.}~\bibnamefont {Zhang}}, \bibinfo {author} {\bibfnamefont {D.}~\bibnamefont {Wang}}, \bibinfo {author} {\bibfnamefont {W.}~\bibnamefont {Chen}}, \bibinfo {author} {\bibfnamefont {W.-Y.}\ \bibnamefont {Yin}}, \ and\ \bibinfo {author} {\bibfnamefont {R.}~\bibnamefont {Zheng}},\ }\bibfield  {title} {\enquote {\bibinfo {title} {Auger scattering dynamic of photo-excited hot carriers in nano-graphite film},}\ }\href {\doibase
  10.1063/5.0116720} {\bibfield  {journal} {\bibinfo  {journal} {Applied Physics Letters}\ }\textbf {\bibinfo {volume} {121}},\ \bibinfo {pages} {181104} (\bibinfo {year} {2022})}\BibitemShut {NoStop}%
\bibitem [{\citenamefont {Lui}\ \emph {et~al.}(1994)\citenamefont {Lui}, \citenamefont {Yamanaka}, \citenamefont {Yoshikuni}, \citenamefont {Seki},\ and\ \citenamefont {Yokoyama}}]{Lui1994}%
  \BibitemOpen
  \bibfield  {author} {\bibinfo {author} {\bibfnamefont {W.~W.}\ \bibnamefont {Lui}}, \bibinfo {author} {\bibfnamefont {T.}~\bibnamefont {Yamanaka}}, \bibinfo {author} {\bibfnamefont {Y.}~\bibnamefont {Yoshikuni}}, \bibinfo {author} {\bibfnamefont {S.}~\bibnamefont {Seki}}, \ and\ \bibinfo {author} {\bibfnamefont {K.}~\bibnamefont {Yokoyama}},\ }\bibfield  {title} {\enquote {\bibinfo {title} {Optimum strain for the suppression of {A}uger recombination effects in compressively strained {InGaAs/InGaAsP} quantum well lasers},}\ }\href {\doibase 10.1063/1.111890} {\bibfield  {journal} {\bibinfo  {journal} {Applied Physics Letters}\ }\textbf {\bibinfo {volume} {64}},\ \bibinfo {pages} {1475--1477} (\bibinfo {year} {1994})}\BibitemShut {NoStop}%
\bibitem [{\citenamefont {Bushick}\ and\ \citenamefont {Kioupakis}(2023)}]{Bushick2023}%
  \BibitemOpen
  \bibfield  {author} {\bibinfo {author} {\bibfnamefont {K.}~\bibnamefont {Bushick}}\ and\ \bibinfo {author} {\bibfnamefont {E.}~\bibnamefont {Kioupakis}},\ }\bibfield  {title} {\enquote {\bibinfo {title} {Phonon-assisted {A}uger-{M}eitner {R}ecombination in {S}ilicon from {F}irst {P}rinciples},}\ }\href {\doibase 10.1103/PhysRevLett.131.076902} {\bibfield  {journal} {\bibinfo  {journal} {Phys. Rev. Lett.}\ }\textbf {\bibinfo {volume} {131}},\ \bibinfo {pages} {076902} (\bibinfo {year} {2023})}\BibitemShut {NoStop}%
\bibitem [{\citenamefont {Giannozzi}\ \emph {et~al.}(2009)\citenamefont {Giannozzi}, \citenamefont {Baroni}, \citenamefont {Bonini}, \citenamefont {Calandra}, \citenamefont {Car}, \citenamefont {Cavazzoni}, \citenamefont {Ceresoli}, \citenamefont {Chiarotti}, \citenamefont {Cococcioni}, \citenamefont {Dabo}, \citenamefont {Corso}, \citenamefont {Gironcoli}, \citenamefont {Fabris}, \citenamefont {Fratesi}, \citenamefont {Gebauer}, \citenamefont {Gerstmann}, \citenamefont {Gougoussis}, \citenamefont {Kokalj}, \citenamefont {Lazzeri}, \citenamefont {Martin-Samos}, \citenamefont {Marzari}, \citenamefont {Mauri}, \citenamefont {Mazzarello}, \citenamefont {Paolini}, \citenamefont {Pasquarello}, \citenamefont {Paulatto}, \citenamefont {Sbraccia}, \citenamefont {Scandolo}, \citenamefont {Sclauzero}, \citenamefont {Seitsonen}, \citenamefont {Smogunov}, \citenamefont {Umari},\ and\ \citenamefont {Wentzcovitch}}]{Giannozzi2009}%
  \BibitemOpen
  \bibfield  {author} {\bibinfo {author} {\bibfnamefont {P.}~\bibnamefont {Giannozzi}}, \bibinfo {author} {\bibfnamefont {S.}~\bibnamefont {Baroni}}, \bibinfo {author} {\bibfnamefont {N.}~\bibnamefont {Bonini}}, \bibinfo {author} {\bibfnamefont {M.}~\bibnamefont {Calandra}}, \bibinfo {author} {\bibfnamefont {R.}~\bibnamefont {Car}}, \bibinfo {author} {\bibfnamefont {C.}~\bibnamefont {Cavazzoni}}, \bibinfo {author} {\bibfnamefont {D.}~\bibnamefont {Ceresoli}}, \bibinfo {author} {\bibfnamefont {G.~L.}\ \bibnamefont {Chiarotti}}, \bibinfo {author} {\bibfnamefont {M.}~\bibnamefont {Cococcioni}}, \bibinfo {author} {\bibfnamefont {I.}~\bibnamefont {Dabo}}, \bibinfo {author} {\bibfnamefont {A.~D.}\ \bibnamefont {Corso}}, \bibinfo {author} {\bibfnamefont {S.~D.}\ \bibnamefont {Gironcoli}}, \bibinfo {author} {\bibfnamefont {S.}~\bibnamefont {Fabris}}, \bibinfo {author} {\bibfnamefont {G.}~\bibnamefont {Fratesi}}, \bibinfo {author} {\bibfnamefont {R.}~\bibnamefont {Gebauer}}, \bibinfo {author} {\bibfnamefont
  {U.}~\bibnamefont {Gerstmann}}, \bibinfo {author} {\bibfnamefont {C.}~\bibnamefont {Gougoussis}}, \bibinfo {author} {\bibfnamefont {A.}~\bibnamefont {Kokalj}}, \bibinfo {author} {\bibfnamefont {M.}~\bibnamefont {Lazzeri}}, \bibinfo {author} {\bibfnamefont {L.}~\bibnamefont {Martin-Samos}}, \bibinfo {author} {\bibfnamefont {N.}~\bibnamefont {Marzari}}, \bibinfo {author} {\bibfnamefont {F.}~\bibnamefont {Mauri}}, \bibinfo {author} {\bibfnamefont {R.}~\bibnamefont {Mazzarello}}, \bibinfo {author} {\bibfnamefont {S.}~\bibnamefont {Paolini}}, \bibinfo {author} {\bibfnamefont {A.}~\bibnamefont {Pasquarello}}, \bibinfo {author} {\bibfnamefont {L.}~\bibnamefont {Paulatto}}, \bibinfo {author} {\bibfnamefont {C.}~\bibnamefont {Sbraccia}}, \bibinfo {author} {\bibfnamefont {S.}~\bibnamefont {Scandolo}}, \bibinfo {author} {\bibfnamefont {G.}~\bibnamefont {Sclauzero}}, \bibinfo {author} {\bibfnamefont {A.~P.}\ \bibnamefont {Seitsonen}}, \bibinfo {author} {\bibfnamefont {A.}~\bibnamefont {Smogunov}}, \bibinfo {author}
  {\bibfnamefont {P.}~\bibnamefont {Umari}}, \ and\ \bibinfo {author} {\bibfnamefont {R.~M.}\ \bibnamefont {Wentzcovitch}},\ }\bibfield  {title} {\enquote {\bibinfo {title} {{QUANTUM ESPRESSO}: {A} modular and open-source software project for quantum simulations of materials},}\ }\href {\doibase 10.1088/0953-8984/21/39/395502} {\bibfield  {journal} {\bibinfo  {journal} {Journal of Physics Condensed Matter}\ }\textbf {\bibinfo {volume} {21}},\ \bibinfo {pages} {395502} (\bibinfo {year} {2009})}\BibitemShut {NoStop}%
\bibitem [{\citenamefont {Giannozzi}\ \emph {et~al.}(2017)\citenamefont {Giannozzi}, \citenamefont {Andreussi}, \citenamefont {Brumme}, \citenamefont {Bunau}, \citenamefont {Nardelli}, \citenamefont {Calandra}, \citenamefont {Car}, \citenamefont {Cavazzoni}, \citenamefont {Ceresoli}, \citenamefont {Cococcioni}, \citenamefont {Colonna}, \citenamefont {Carnimeo}, \citenamefont {Corso}, \citenamefont {de~Gironcoli}, \citenamefont {Delugas}, \citenamefont {DiStasio}, \citenamefont {Ferretti}, \citenamefont {Floris}, \citenamefont {Fratesi}, \citenamefont {Fugallo}, \citenamefont {Gebauer}, \citenamefont {Gerstmann}, \citenamefont {Giustino}, \citenamefont {Gorni}, \citenamefont {Jia}, \citenamefont {Kawamura}, \citenamefont {Ko}, \citenamefont {Kokalj}, \citenamefont {Küçükbenli}, \citenamefont {Lazzeri}, \citenamefont {Marsili}, \citenamefont {Marzari}, \citenamefont {Mauri}, \citenamefont {Nguyen}, \citenamefont {Nguyen}, \citenamefont {de-la Roza}, \citenamefont {Paulatto}, \citenamefont {Poncé},
  \citenamefont {Rocca}, \citenamefont {Sabatini}, \citenamefont {Santra}, \citenamefont {Schlipf}, \citenamefont {Seitsonen}, \citenamefont {Smogunov}, \citenamefont {Timrov}, \citenamefont {Thonhauser}, \citenamefont {Umari}, \citenamefont {Vast}, \citenamefont {Wu},\ and\ \citenamefont {Baroni}}]{Giannozzi2017}%
  \BibitemOpen
  \bibfield  {author} {\bibinfo {author} {\bibfnamefont {P.}~\bibnamefont {Giannozzi}}, \bibinfo {author} {\bibfnamefont {O.}~\bibnamefont {Andreussi}}, \bibinfo {author} {\bibfnamefont {T.}~\bibnamefont {Brumme}}, \bibinfo {author} {\bibfnamefont {O.}~\bibnamefont {Bunau}}, \bibinfo {author} {\bibfnamefont {M.~B.}\ \bibnamefont {Nardelli}}, \bibinfo {author} {\bibfnamefont {M.}~\bibnamefont {Calandra}}, \bibinfo {author} {\bibfnamefont {R.}~\bibnamefont {Car}}, \bibinfo {author} {\bibfnamefont {C.}~\bibnamefont {Cavazzoni}}, \bibinfo {author} {\bibfnamefont {D.}~\bibnamefont {Ceresoli}}, \bibinfo {author} {\bibfnamefont {M.}~\bibnamefont {Cococcioni}}, \bibinfo {author} {\bibfnamefont {N.}~\bibnamefont {Colonna}}, \bibinfo {author} {\bibfnamefont {I.}~\bibnamefont {Carnimeo}}, \bibinfo {author} {\bibfnamefont {A.~D.}\ \bibnamefont {Corso}}, \bibinfo {author} {\bibfnamefont {S.}~\bibnamefont {de~Gironcoli}}, \bibinfo {author} {\bibfnamefont {P.}~\bibnamefont {Delugas}}, \bibinfo {author} {\bibfnamefont
  {R.~A.}\ \bibnamefont {DiStasio}}, \bibinfo {author} {\bibfnamefont {A.}~\bibnamefont {Ferretti}}, \bibinfo {author} {\bibfnamefont {A.}~\bibnamefont {Floris}}, \bibinfo {author} {\bibfnamefont {G.}~\bibnamefont {Fratesi}}, \bibinfo {author} {\bibfnamefont {G.}~\bibnamefont {Fugallo}}, \bibinfo {author} {\bibfnamefont {R.}~\bibnamefont {Gebauer}}, \bibinfo {author} {\bibfnamefont {U.}~\bibnamefont {Gerstmann}}, \bibinfo {author} {\bibfnamefont {F.}~\bibnamefont {Giustino}}, \bibinfo {author} {\bibfnamefont {T.}~\bibnamefont {Gorni}}, \bibinfo {author} {\bibfnamefont {J.}~\bibnamefont {Jia}}, \bibinfo {author} {\bibfnamefont {M.}~\bibnamefont {Kawamura}}, \bibinfo {author} {\bibfnamefont {H.-Y.}\ \bibnamefont {Ko}}, \bibinfo {author} {\bibfnamefont {A.}~\bibnamefont {Kokalj}}, \bibinfo {author} {\bibfnamefont {E.}~\bibnamefont {Küçükbenli}}, \bibinfo {author} {\bibfnamefont {M.}~\bibnamefont {Lazzeri}}, \bibinfo {author} {\bibfnamefont {M.}~\bibnamefont {Marsili}}, \bibinfo {author} {\bibfnamefont
  {N.}~\bibnamefont {Marzari}}, \bibinfo {author} {\bibfnamefont {F.}~\bibnamefont {Mauri}}, \bibinfo {author} {\bibfnamefont {N.~L.}\ \bibnamefont {Nguyen}}, \bibinfo {author} {\bibfnamefont {H.-V.}\ \bibnamefont {Nguyen}}, \bibinfo {author} {\bibfnamefont {A.~O.}\ \bibnamefont {de-la Roza}}, \bibinfo {author} {\bibfnamefont {L.}~\bibnamefont {Paulatto}}, \bibinfo {author} {\bibfnamefont {S.}~\bibnamefont {Poncé}}, \bibinfo {author} {\bibfnamefont {D.}~\bibnamefont {Rocca}}, \bibinfo {author} {\bibfnamefont {R.}~\bibnamefont {Sabatini}}, \bibinfo {author} {\bibfnamefont {B.}~\bibnamefont {Santra}}, \bibinfo {author} {\bibfnamefont {M.}~\bibnamefont {Schlipf}}, \bibinfo {author} {\bibfnamefont {A.~P.}\ \bibnamefont {Seitsonen}}, \bibinfo {author} {\bibfnamefont {A.}~\bibnamefont {Smogunov}}, \bibinfo {author} {\bibfnamefont {I.}~\bibnamefont {Timrov}}, \bibinfo {author} {\bibfnamefont {T.}~\bibnamefont {Thonhauser}}, \bibinfo {author} {\bibfnamefont {P.}~\bibnamefont {Umari}}, \bibinfo {author}
  {\bibfnamefont {N.}~\bibnamefont {Vast}}, \bibinfo {author} {\bibfnamefont {X.}~\bibnamefont {Wu}}, \ and\ \bibinfo {author} {\bibfnamefont {S.}~\bibnamefont {Baroni}},\ }\bibfield  {title} {\enquote {\bibinfo {title} {Advanced capabilities for materials modelling with {Q}uantum {ESPRESSO}},}\ }\href {\doibase 10.1088/1361-648X/aa8f79} {\bibfield  {journal} {\bibinfo  {journal} {Journal of Physics: Condensed Matter}\ }\textbf {\bibinfo {volume} {29}},\ \bibinfo {pages} {465901} (\bibinfo {year} {2017})}\BibitemShut {NoStop}%
\bibitem [{\citenamefont {Giannozzi}\ \emph {et~al.}(2020)\citenamefont {Giannozzi}, \citenamefont {Baseggio}, \citenamefont {Bonfà}, \citenamefont {Brunato}, \citenamefont {Car}, \citenamefont {Carnimeo}, \citenamefont {Cavazzoni}, \citenamefont {de~Gironcoli}, \citenamefont {Delugas}, \citenamefont {Ruffino}, \citenamefont {Ferretti}, \citenamefont {Marzari}, \citenamefont {Timrov}, \citenamefont {Urru},\ and\ \citenamefont {Baroni}}]{Giannozzi2020}%
  \BibitemOpen
  \bibfield  {author} {\bibinfo {author} {\bibfnamefont {P.}~\bibnamefont {Giannozzi}}, \bibinfo {author} {\bibfnamefont {O.}~\bibnamefont {Baseggio}}, \bibinfo {author} {\bibfnamefont {P.}~\bibnamefont {Bonfà}}, \bibinfo {author} {\bibfnamefont {D.}~\bibnamefont {Brunato}}, \bibinfo {author} {\bibfnamefont {R.}~\bibnamefont {Car}}, \bibinfo {author} {\bibfnamefont {I.}~\bibnamefont {Carnimeo}}, \bibinfo {author} {\bibfnamefont {C.}~\bibnamefont {Cavazzoni}}, \bibinfo {author} {\bibfnamefont {S.}~\bibnamefont {de~Gironcoli}}, \bibinfo {author} {\bibfnamefont {P.}~\bibnamefont {Delugas}}, \bibinfo {author} {\bibfnamefont {F.~F.}\ \bibnamefont {Ruffino}}, \bibinfo {author} {\bibfnamefont {A.}~\bibnamefont {Ferretti}}, \bibinfo {author} {\bibfnamefont {N.}~\bibnamefont {Marzari}}, \bibinfo {author} {\bibfnamefont {I.}~\bibnamefont {Timrov}}, \bibinfo {author} {\bibfnamefont {A.}~\bibnamefont {Urru}}, \ and\ \bibinfo {author} {\bibfnamefont {S.}~\bibnamefont {Baroni}},\ }\bibfield  {title} {\enquote {\bibinfo
  {title} {Quantum {ESPRESSO} toward the exascale},}\ }\href {\doibase 10.1063/5.0005082} {\bibfield  {journal} {\bibinfo  {journal} {The Journal of Chemical Physics}\ }\textbf {\bibinfo {volume} {152}},\ \bibinfo {pages} {154105} (\bibinfo {year} {2020})}\BibitemShut {NoStop}%
\bibitem [{\citenamefont {Ceperley}\ and\ \citenamefont {Alder}(1980)}]{Ceperley1980}%
  \BibitemOpen
  \bibfield  {author} {\bibinfo {author} {\bibfnamefont {D.~M.}\ \bibnamefont {Ceperley}}\ and\ \bibinfo {author} {\bibfnamefont {B.~J.}\ \bibnamefont {Alder}},\ }\bibfield  {title} {\enquote {\bibinfo {title} {Ground {S}tate of the {E}lectron {G}as by a {S}tochastic {M}ethod},}\ }\href {\doibase 10.1103/PhysRevLett.45.566} {\bibfield  {journal} {\bibinfo  {journal} {Phys. Rev. Lett.}\ }\textbf {\bibinfo {volume} {45}},\ \bibinfo {pages} {566--569} (\bibinfo {year} {1980})}\BibitemShut {NoStop}%
\bibitem [{\citenamefont {Perdew}\ and\ \citenamefont {Zunger}(1981)}]{Perdew1981}%
  \BibitemOpen
  \bibfield  {author} {\bibinfo {author} {\bibfnamefont {J.~P.}\ \bibnamefont {Perdew}}\ and\ \bibinfo {author} {\bibfnamefont {A.}~\bibnamefont {Zunger}},\ }\bibfield  {title} {\enquote {\bibinfo {title} {Self-interaction correction to density-functional approximations for many-electron systems},}\ }\href {\doibase 10.1103/PhysRevB.23.5048} {\bibfield  {journal} {\bibinfo  {journal} {Phys. Rev. B}\ }\textbf {\bibinfo {volume} {23}},\ \bibinfo {pages} {5048--5079} (\bibinfo {year} {1981})}\BibitemShut {NoStop}%
\bibitem [{\citenamefont {Troullier}\ and\ \citenamefont {Martins}(1991)}]{Troullier1991}%
  \BibitemOpen
  \bibfield  {author} {\bibinfo {author} {\bibfnamefont {N.}~\bibnamefont {Troullier}}\ and\ \bibinfo {author} {\bibfnamefont {J.~L.}\ \bibnamefont {Martins}},\ }\bibfield  {title} {\enquote {\bibinfo {title} {Efficient pseudopotentials for plane-wave calculations},}\ }\href {\doibase 10.1103/PhysRevB.43.1993} {\bibfield  {journal} {\bibinfo  {journal} {Phys. Rev. B}\ }\textbf {\bibinfo {volume} {43}},\ \bibinfo {pages} {1993--2006} (\bibinfo {year} {1991})}\BibitemShut {NoStop}%
\bibitem [{\citenamefont {Bona}\ and\ \citenamefont {Meier}(1985)}]{Bona1985}%
  \BibitemOpen
  \bibfield  {author} {\bibinfo {author} {\bibfnamefont {G.}~\bibnamefont {Bona}}\ and\ \bibinfo {author} {\bibfnamefont {F.}~\bibnamefont {Meier}},\ }\bibfield  {title} {\enquote {\bibinfo {title} {Observation of the spin-orbit splitting at the valence band edge of silicon by spin-polarized photoemission},}\ }\href {\doibase https://doi.org/10.1016/0038-1098(85)90813-0} {\bibfield  {journal} {\bibinfo  {journal} {Solid State Communications}\ }\textbf {\bibinfo {volume} {55}},\ \bibinfo {pages} {851--855} (\bibinfo {year} {1985})}\BibitemShut {NoStop}%
\bibitem [{\citenamefont {Ponc\'e}\ \emph {et~al.}(2021)\citenamefont {Ponc\'e}, \citenamefont {Macheda}, \citenamefont {Margine}, \citenamefont {Marzari}, \citenamefont {Bonini},\ and\ \citenamefont {Giustino}}]{Ponce2021}%
  \BibitemOpen
  \bibfield  {author} {\bibinfo {author} {\bibfnamefont {S.}~\bibnamefont {Ponc\'e}}, \bibinfo {author} {\bibfnamefont {F.}~\bibnamefont {Macheda}}, \bibinfo {author} {\bibfnamefont {E.~R.}\ \bibnamefont {Margine}}, \bibinfo {author} {\bibfnamefont {N.}~\bibnamefont {Marzari}}, \bibinfo {author} {\bibfnamefont {N.}~\bibnamefont {Bonini}}, \ and\ \bibinfo {author} {\bibfnamefont {F.}~\bibnamefont {Giustino}},\ }\bibfield  {title} {\enquote {\bibinfo {title} {First-principles predictions of {H}all and drift mobilities in semiconductors},}\ }\href {\doibase 10.1103/PhysRevResearch.3.043022} {\bibfield  {journal} {\bibinfo  {journal} {Phys. Rev. Res.}\ }\textbf {\bibinfo {volume} {3}},\ \bibinfo {pages} {043022} (\bibinfo {year} {2021})}\BibitemShut {NoStop}%
\bibitem [{\citenamefont {Zacharias}, \citenamefont {Scheffler},\ and\ \citenamefont {Carbogno}(2020)}]{Zacharias2020}%
  \BibitemOpen
  \bibfield  {author} {\bibinfo {author} {\bibfnamefont {M.}~\bibnamefont {Zacharias}}, \bibinfo {author} {\bibfnamefont {M.}~\bibnamefont {Scheffler}}, \ and\ \bibinfo {author} {\bibfnamefont {C.}~\bibnamefont {Carbogno}},\ }\bibfield  {title} {\enquote {\bibinfo {title} {Fully anharmonic nonperturbative theory of vibronically renormalized electronic band structures},}\ }\href {\doibase 10.1103/PhysRevB.102.045126} {\bibfield  {journal} {\bibinfo  {journal} {Phys. Rev. B}\ }\textbf {\bibinfo {volume} {102}},\ \bibinfo {pages} {045126} (\bibinfo {year} {2020})}\BibitemShut {NoStop}%
\bibitem [{\citenamefont {Marzari}\ \emph {et~al.}(2012)\citenamefont {Marzari}, \citenamefont {Mostofi}, \citenamefont {Yates}, \citenamefont {Souza},\ and\ \citenamefont {Vanderbilt}}]{Marzari2012}%
  \BibitemOpen
  \bibfield  {author} {\bibinfo {author} {\bibfnamefont {N.}~\bibnamefont {Marzari}}, \bibinfo {author} {\bibfnamefont {A.~A.}\ \bibnamefont {Mostofi}}, \bibinfo {author} {\bibfnamefont {J.~R.}\ \bibnamefont {Yates}}, \bibinfo {author} {\bibfnamefont {I.}~\bibnamefont {Souza}}, \ and\ \bibinfo {author} {\bibfnamefont {D.}~\bibnamefont {Vanderbilt}},\ }\bibfield  {title} {\enquote {\bibinfo {title} {Maximally localized {W}annier functions: {T}heory and applications},}\ }\href {\doibase 10.1103/RevModPhys.84.1419} {\bibfield  {journal} {\bibinfo  {journal} {Rev. Mod. Phys.}\ }\textbf {\bibinfo {volume} {84}},\ \bibinfo {pages} {1419--1475} (\bibinfo {year} {2012})}\BibitemShut {NoStop}%
\bibitem [{\citenamefont {Pizzi}\ \emph {et~al.}(2020)\citenamefont {Pizzi}, \citenamefont {Vitale}, \citenamefont {Arita}, \citenamefont {Bl{\"{u}}gel}, \citenamefont {Freimuth}, \citenamefont {G{\'{e}}ranton}, \citenamefont {Gibertini}, \citenamefont {Gresch}, \citenamefont {Johnson}, \citenamefont {Koretsune}, \citenamefont {Iba{\~{n}}ez-Azpiroz}, \citenamefont {Lee}, \citenamefont {Lihm}, \citenamefont {Marchand}, \citenamefont {Marrazzo}, \citenamefont {Mokrousov}, \citenamefont {Mustafa}, \citenamefont {Nohara}, \citenamefont {Nomura}, \citenamefont {Paulatto}, \citenamefont {Ponc{\'{e}}}, \citenamefont {Ponweiser}, \citenamefont {Qiao}, \citenamefont {Th{\"{o}}le}, \citenamefont {Tsirkin}, \citenamefont {Wierzbowska}, \citenamefont {Marzari}, \citenamefont {Vanderbilt}, \citenamefont {Souza}, \citenamefont {Mostofi},\ and\ \citenamefont {Yates}}]{Pizzi2020}%
  \BibitemOpen
  \bibfield  {author} {\bibinfo {author} {\bibfnamefont {G.}~\bibnamefont {Pizzi}}, \bibinfo {author} {\bibfnamefont {V.}~\bibnamefont {Vitale}}, \bibinfo {author} {\bibfnamefont {R.}~\bibnamefont {Arita}}, \bibinfo {author} {\bibfnamefont {S.}~\bibnamefont {Bl{\"{u}}gel}}, \bibinfo {author} {\bibfnamefont {F.}~\bibnamefont {Freimuth}}, \bibinfo {author} {\bibfnamefont {G.}~\bibnamefont {G{\'{e}}ranton}}, \bibinfo {author} {\bibfnamefont {M.}~\bibnamefont {Gibertini}}, \bibinfo {author} {\bibfnamefont {D.}~\bibnamefont {Gresch}}, \bibinfo {author} {\bibfnamefont {C.}~\bibnamefont {Johnson}}, \bibinfo {author} {\bibfnamefont {T.}~\bibnamefont {Koretsune}}, \bibinfo {author} {\bibfnamefont {J.}~\bibnamefont {Iba{\~{n}}ez-Azpiroz}}, \bibinfo {author} {\bibfnamefont {H.}~\bibnamefont {Lee}}, \bibinfo {author} {\bibfnamefont {J.-M.}\ \bibnamefont {Lihm}}, \bibinfo {author} {\bibfnamefont {D.}~\bibnamefont {Marchand}}, \bibinfo {author} {\bibfnamefont {A.}~\bibnamefont {Marrazzo}}, \bibinfo {author} {\bibfnamefont
  {Y.}~\bibnamefont {Mokrousov}}, \bibinfo {author} {\bibfnamefont {J.~I.}\ \bibnamefont {Mustafa}}, \bibinfo {author} {\bibfnamefont {Y.}~\bibnamefont {Nohara}}, \bibinfo {author} {\bibfnamefont {Y.}~\bibnamefont {Nomura}}, \bibinfo {author} {\bibfnamefont {L.}~\bibnamefont {Paulatto}}, \bibinfo {author} {\bibfnamefont {S.}~\bibnamefont {Ponc{\'{e}}}}, \bibinfo {author} {\bibfnamefont {T.}~\bibnamefont {Ponweiser}}, \bibinfo {author} {\bibfnamefont {J.}~\bibnamefont {Qiao}}, \bibinfo {author} {\bibfnamefont {F.}~\bibnamefont {Th{\"{o}}le}}, \bibinfo {author} {\bibfnamefont {S.~S.}\ \bibnamefont {Tsirkin}}, \bibinfo {author} {\bibfnamefont {M.}~\bibnamefont {Wierzbowska}}, \bibinfo {author} {\bibfnamefont {N.}~\bibnamefont {Marzari}}, \bibinfo {author} {\bibfnamefont {D.}~\bibnamefont {Vanderbilt}}, \bibinfo {author} {\bibfnamefont {I.}~\bibnamefont {Souza}}, \bibinfo {author} {\bibfnamefont {A.~A.}\ \bibnamefont {Mostofi}}, \ and\ \bibinfo {author} {\bibfnamefont {J.~R.}\ \bibnamefont {Yates}},\ }\bibfield
  {title} {\enquote {\bibinfo {title} {{Wannier90 as a community code: new features and applications}},}\ }\href {\doibase 10.1088/1361-648X/ab51ff} {\bibfield  {journal} {\bibinfo  {journal} {Journal of Physics: Condensed Matter}\ }\textbf {\bibinfo {volume} {32}},\ \bibinfo {pages} {165902} (\bibinfo {year} {2020})}\BibitemShut {NoStop}%
\bibitem [{\citenamefont {Bludau}, \citenamefont {Onton},\ and\ \citenamefont {Heinke}(1974)}]{Bludau1974}%
  \BibitemOpen
  \bibfield  {author} {\bibinfo {author} {\bibfnamefont {W.}~\bibnamefont {Bludau}}, \bibinfo {author} {\bibfnamefont {A.}~\bibnamefont {Onton}}, \ and\ \bibinfo {author} {\bibfnamefont {W.}~\bibnamefont {Heinke}},\ }\bibfield  {title} {\enquote {\bibinfo {title} {{Temperature dependence of the band gap of silicon}},}\ }\href {\doibase 10.1063/1.1663501} {\bibfield  {journal} {\bibinfo  {journal} {Journal of Applied Physics}\ }\textbf {\bibinfo {volume} {45}},\ \bibinfo {pages} {1846--1848} (\bibinfo {year} {1974})}\BibitemShut {NoStop}%
\bibitem [{\citenamefont {Munguía}\ \emph {et~al.}(2008)\citenamefont {Munguía}, \citenamefont {Bremond}, \citenamefont {Bluet}, \citenamefont {Hartmann},\ and\ \citenamefont {Mermoux}}]{Munguia2008}%
  \BibitemOpen
  \bibfield  {author} {\bibinfo {author} {\bibfnamefont {J.}~\bibnamefont {Munguía}}, \bibinfo {author} {\bibfnamefont {G.}~\bibnamefont {Bremond}}, \bibinfo {author} {\bibfnamefont {J.~M.}\ \bibnamefont {Bluet}}, \bibinfo {author} {\bibfnamefont {J.~M.}\ \bibnamefont {Hartmann}}, \ and\ \bibinfo {author} {\bibfnamefont {M.}~\bibnamefont {Mermoux}},\ }\bibfield  {title} {\enquote {\bibinfo {title} {Strain dependence of indirect band gap for strained silicon on insulator wafers},}\ }\href {\doibase 10.1063/1.2978241} {\bibfield  {journal} {\bibinfo  {journal} {Applied Physics Letters}\ }\textbf {\bibinfo {volume} {93}},\ \bibinfo {pages} {102101} (\bibinfo {year} {2008})}\BibitemShut {NoStop}%
\bibitem [{\citenamefont {Richter}\ \emph {et~al.}(2012)\citenamefont {Richter}, \citenamefont {Glunz}, \citenamefont {Werner}, \citenamefont {Schmidt},\ and\ \citenamefont {Cuevas}}]{Richter2012}%
  \BibitemOpen
  \bibfield  {author} {\bibinfo {author} {\bibfnamefont {A.}~\bibnamefont {Richter}}, \bibinfo {author} {\bibfnamefont {S.~W.}\ \bibnamefont {Glunz}}, \bibinfo {author} {\bibfnamefont {F.}~\bibnamefont {Werner}}, \bibinfo {author} {\bibfnamefont {J.}~\bibnamefont {Schmidt}}, \ and\ \bibinfo {author} {\bibfnamefont {A.}~\bibnamefont {Cuevas}},\ }\bibfield  {title} {\enquote {\bibinfo {title} {Improved quantitative description of {A}uger recombination in crystalline silicon},}\ }\href {\doibase 10.1103/PhysRevB.86.165202} {\bibfield  {journal} {\bibinfo  {journal} {Phys. Rev. B}\ }\textbf {\bibinfo {volume} {86}},\ \bibinfo {pages} {165202} (\bibinfo {year} {2012})}\BibitemShut {NoStop}%
\bibitem [{\citenamefont {Kioupakis}\ \emph {et~al.}(2015)\citenamefont {Kioupakis}, \citenamefont {Steiauf}, \citenamefont {Rinke}, \citenamefont {Delaney},\ and\ \citenamefont {Walle}}]{Kioupakis2015}%
  \BibitemOpen
  \bibfield  {author} {\bibinfo {author} {\bibfnamefont {E.}~\bibnamefont {Kioupakis}}, \bibinfo {author} {\bibfnamefont {D.}~\bibnamefont {Steiauf}}, \bibinfo {author} {\bibfnamefont {P.}~\bibnamefont {Rinke}}, \bibinfo {author} {\bibfnamefont {K.~T.}\ \bibnamefont {Delaney}}, \ and\ \bibinfo {author} {\bibfnamefont {C.~G. V.~D.}\ \bibnamefont {Walle}},\ }\bibfield  {title} {\enquote {\bibinfo {title} {First-principles calculations of indirect {A}uger recombination in nitride semiconductors},}\ }\href {\doibase 10.1103/PhysRevB.92.035207} {\bibfield  {journal} {\bibinfo  {journal} {Physical Review B}\ }\textbf {\bibinfo {volume} {92}},\ \bibinfo {pages} {035207} (\bibinfo {year} {2015})}\BibitemShut {NoStop}%
\bibitem [{\citenamefont {Dziewior}\ and\ \citenamefont {Schmid}(1977)}]{Dziewior1977}%
  \BibitemOpen
  \bibfield  {author} {\bibinfo {author} {\bibfnamefont {J.}~\bibnamefont {Dziewior}}\ and\ \bibinfo {author} {\bibfnamefont {W.}~\bibnamefont {Schmid}},\ }\bibfield  {title} {\enquote {\bibinfo {title} {Auger coefficients for highly doped and highly excited silicon},}\ }\href {\doibase 10.1063/1.89694} {\bibfield  {journal} {\bibinfo  {journal} {Applied Physics Letters}\ }\textbf {\bibinfo {volume} {31}},\ \bibinfo {pages} {346--348} (\bibinfo {year} {1977})}\BibitemShut {NoStop}%
\bibitem [{\citenamefont {Häcker}\ and\ \citenamefont {Hangleiter}(1994)}]{Hacker1994}%
  \BibitemOpen
  \bibfield  {author} {\bibinfo {author} {\bibfnamefont {R.}~\bibnamefont {Häcker}}\ and\ \bibinfo {author} {\bibfnamefont {A.}~\bibnamefont {Hangleiter}},\ }\bibfield  {title} {\enquote {\bibinfo {title} {Intrinsic upper limits of the carrier lifetime in silicon},}\ }\href {\doibase 10.1063/1.356634} {\bibfield  {journal} {\bibinfo  {journal} {Journal of Applied Physics}\ }\textbf {\bibinfo {volume} {75}},\ \bibinfo {pages} {7570--7572} (\bibinfo {year} {1994})}\BibitemShut {NoStop}%
\bibitem [{\citenamefont {Govoni}, \citenamefont {Marri},\ and\ \citenamefont {Ossicini}(2011)}]{Govoni2011}%
  \BibitemOpen
  \bibfield  {author} {\bibinfo {author} {\bibfnamefont {M.}~\bibnamefont {Govoni}}, \bibinfo {author} {\bibfnamefont {I.}~\bibnamefont {Marri}}, \ and\ \bibinfo {author} {\bibfnamefont {S.}~\bibnamefont {Ossicini}},\ }\bibfield  {title} {\enquote {\bibinfo {title} {Auger recombination in {S}i and {G}a{A}s semiconductors: {A}b initio results},}\ }\href {\doibase 10.1103/PhysRevB.84.075215} {\bibfield  {journal} {\bibinfo  {journal} {Physical Review B}\ }\textbf {\bibinfo {volume} {84}},\ \bibinfo {pages} {075215} (\bibinfo {year} {2011})}\BibitemShut {NoStop}%
\bibitem [{\citenamefont {Dominici}\ \emph {et~al.}(2016)\citenamefont {Dominici}, \citenamefont {Wen}, \citenamefont {Bertazzi}, \citenamefont {Goano},\ and\ \citenamefont {Bellotti}}]{Dominici2016}%
  \BibitemOpen
  \bibfield  {author} {\bibinfo {author} {\bibfnamefont {S.}~\bibnamefont {Dominici}}, \bibinfo {author} {\bibfnamefont {H.}~\bibnamefont {Wen}}, \bibinfo {author} {\bibfnamefont {F.}~\bibnamefont {Bertazzi}}, \bibinfo {author} {\bibfnamefont {M.}~\bibnamefont {Goano}}, \ and\ \bibinfo {author} {\bibfnamefont {E.}~\bibnamefont {Bellotti}},\ }\bibfield  {title} {\enquote {\bibinfo {title} {{Numerical evaluation of Auger recombination coefficients in relaxed and strained germanium}},}\ }\href {\doibase 10.1063/1.4952720} {\bibfield  {journal} {\bibinfo  {journal} {Applied Physics Letters}\ }\textbf {\bibinfo {volume} {108}},\ \bibinfo {pages} {211103} (\bibinfo {year} {2016})}\BibitemShut {NoStop}%
\bibitem [{\citenamefont {Sze}\ and\ \citenamefont {Ng}(2006)}]{Sze2006}%
  \BibitemOpen
  \bibfield  {author} {\bibinfo {author} {\bibfnamefont {S.}~\bibnamefont {Sze}}\ and\ \bibinfo {author} {\bibfnamefont {K.~K.}\ \bibnamefont {Ng}},\ }\href {\doibase 10.1002/0470068329} {\emph {\bibinfo {title} {Physics of {S}emiconductor {D}evices}}}\ (\bibinfo  {publisher} {John Wiley \& Sons, Inc.},\ \bibinfo {year} {2006})\BibitemShut {NoStop}%
\bibitem [{\citenamefont {Yu}, \citenamefont {Zhang},\ and\ \citenamefont {Liu}(2008)}]{Yu2008}%
  \BibitemOpen
  \bibfield  {author} {\bibinfo {author} {\bibfnamefont {D.}~\bibnamefont {Yu}}, \bibinfo {author} {\bibfnamefont {Y.}~\bibnamefont {Zhang}}, \ and\ \bibinfo {author} {\bibfnamefont {F.}~\bibnamefont {Liu}},\ }\bibfield  {title} {\enquote {\bibinfo {title} {First-principles study of electronic properties of biaxially strained silicon: Effects on charge carrier mobility},}\ }\href {\doibase 10.1103/PhysRevB.78.245204} {\bibfield  {journal} {\bibinfo  {journal} {Physical Review B}\ }\textbf {\bibinfo {volume} {78}},\ \bibinfo {pages} {245204} (\bibinfo {year} {2008})}\BibitemShut {NoStop}%
\end{thebibliography}%

\clearpage
\beginsupplement
\appendix

\begin{center}
    \large{\textbf{Supplementary Material for Strain Effects on Auger-Meitner Recombination in Silicon}}
    
    \vspace{3mm}
    
    \normalsize{Kyle Bushick\textsuperscript{1,2} and Emmanouil Kioupakis\textsuperscript{1}}
    
    \small{\textit{\textsuperscript{1)} Department of Materials Science and Engineering, University of Michigan, Ann Arbor, MI, USA}}
    
    \small{\textit{\textsuperscript{2)} Materials Science Division, Lawrence Livermore National Laboratory, Livermore, CA, USA}}
\end{center}
\bigskip

\begin{center}
\textbf{Strain Effects on Silicon Phonon Dispersion}
\end{center}
While the application of 1\% biaxial strain leads to significant qualitative differences in the electronic band structure, the effects on the phonon dispersion are less notable, and modify the phonon energies by at most \(\sim\)1.2 meV, though for the lower-energy modes, which dominate the phonon-assisted AMR process, the splitting is about half as large. These shifts in the phonon energies are consistent with the shifts of the peak in Fig. 4(a-c), which are at 17.2, 18.2, and 18.4 meV for the compressive, unstrained, and tensile conditions, respectively. As a result, it affects the dominant phonon occupancies (and hence the corresponding phonon-assisted AMR coefficients) by no more than 9\%.

\begin{figure}[ht]
\includegraphics{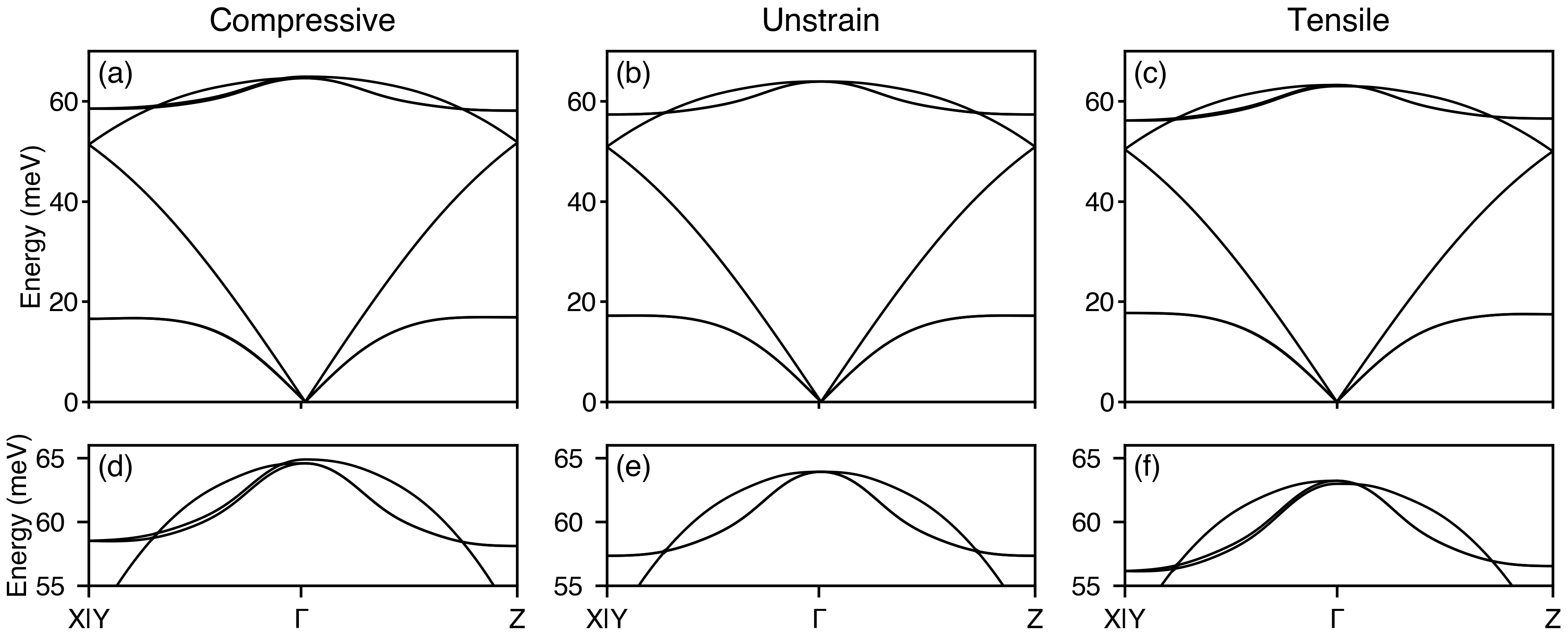}
\caption{Phonon dispersion along high-symmetry paths for silicon under (a) compressive, (b) unstrained, and (c) tensile strain conditions. The strain induced symmetry breaking leads to small variations (\(<2\) meV) to the phonon energies. Plots (d-f) show the details of the optical branches, where the effects of strain are most prominent.}
\label{fig:phonons}
\end{figure}

While the phonon splitting is small, the lifting of degeneracies for both electron bands and phonon modes will affect the electron-phonon matrix elements. We expect that the electron-phonon matrix elements will increase between states of the same symmetry while being suppressed between states of different symmetries. This effect is combined with the symmetry-breaking effects on the Coulomb matrix elements to yield the variations to the AMR coefficients in Figs. 2 and 3.

\newpage

\begin{figure}[ht]
\includegraphics{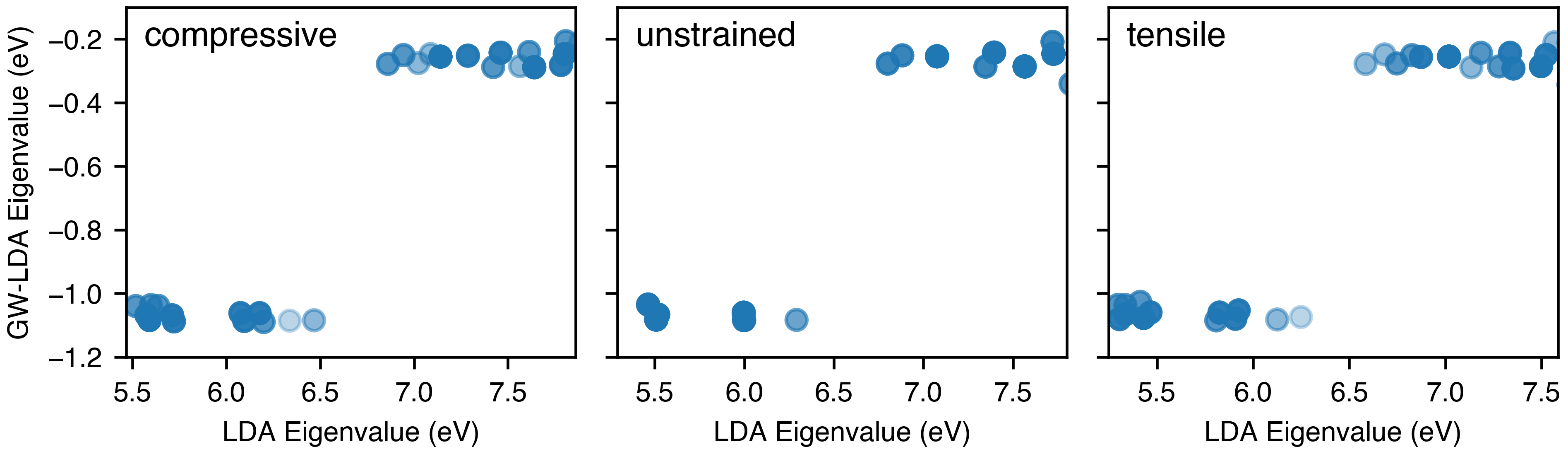}
\caption{Difference between the GW quasiparticle energies and the LDA eigenvalues as a function of LDA eigenvalue energy shown within 1 eV from the corresponding band edges for (a) compressive, (b) unstrained, and (c) tensile strain conditions.}
\label{fig:gw_scissor}
\end{figure}

\begin{center}
\textbf{GW Quasiparticle Corrections}
\end{center}
We compute the GW corrections to the local density approximation (LDA) band-structure eigenvalues under all strain conditions tested. We perform these calculations on an \(8\times8\times8\) Brillouin-zone (BZ) sampling grid with a 35 Ry dielectric-matrix cutoff and a 30 Ry screened Coulomb cutoff. As we show in Fig. \ref{fig:gw_scissor}, the scissor correction slope is effectively zero for all bands, and therefore the effects of GW corrections can be readily approximated with a rigid shift of the eigenvalues across all strain conditions.  

\newpage

\begin{figure}[ht]
\includegraphics{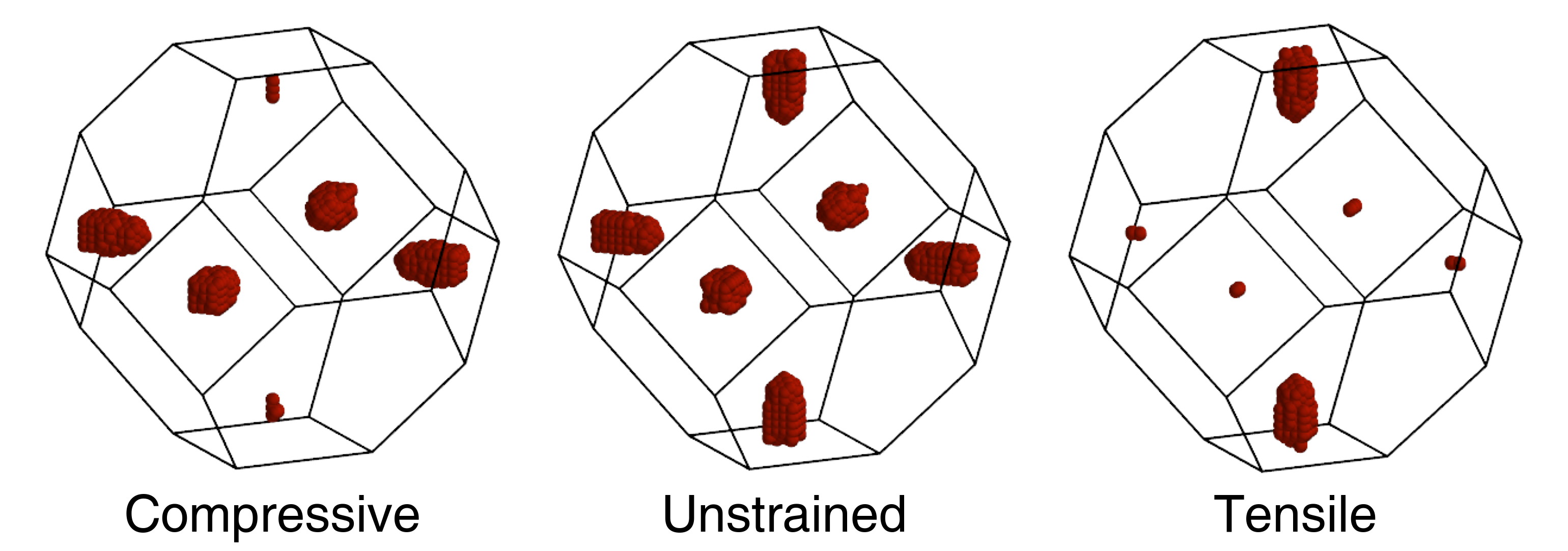}
\caption{Occupation/sampling of conduction band valleys under strain conditions. The low-energy valleys account for the vast majority of contributing electronic states.}
\label{fig:elec_dist}
\end{figure}

\begin{center}
\textbf{Initial Electron Distribution for Direct AMR}
\end{center}
Under the \(\pm1\%\) strain conditions we tested, the valley splitting is strong enough that the vast majority of electrons occupy the low-energy valleys. Figure \ref{fig:elec_dist} shows the visual enumeration of the BZ sampling \textbf{k}-points that fall within our selected energy cutoff {\textendash} within approximately 200 meV of the conduction band minimum {\textendash} accounting for 99.9\% of the charge density. Specifically, 796 of 804 sampling \textbf{k}-points in the compressive system and 345 of 354 sampling \textbf{k}-points in the tensile system belonging to the low-energy valleys. Therefore, the assumption we make for the \pa calculations {\textendash} that the periodic part of the wave functions of all electrons can be approximated by that at the extremum of the low-energy corresponding valley {\textendash} is justified.

\begin{center}
\textbf{Valley Analysis for \(eeh\) AMR}
\end{center}
We extend our valley analysis of the \(eeh\) process in unstrained silicon\cite{Bushick2023} to the strained conditions. We follow the same naming convention used for phonon-scattering processes to delineate the different valley arrangements of the initial states of the two low-energy electrons (also shown visually in the inset of Fig. 3b of the main text): the intravalley process (blue) corresponds to both electrons originating in the same conduction band valley, the \textit{g}-type process (orange) is for electrons that originate in opposite conduction band valleys, and finally in the \textit{f}-type process (green) the electrons originate in perpendicular valleys. As we show in Fig. 3 of the main text, for both the direct and \pa \(eeh\) processes, the \textit{f}-type recombination {\textendash} involving electrons from perpendicular valleys {\textendash} is effectively eliminated for the tensile-strain condition despite the fact that it is the dominant pathway in the unstrained crystal. While this is the most drastic example of strain effects, we also observe a notable increase in the intravalley process under both compressive and tensile strain conditions. For direct \(eeh\), this increase is approximately proportional to the electron concentration increase in each valley (\(1.28\times\) for compressive and \(3.6\times\) for tensile). These increases are approximately \(3\times\) larger for the \pa process, with scaling of \(4.23\times\) and \(8.59\times\) for the compressive and tensile cases, respectively. Overall, the total \(eeh\) AMR coefficient under strain conditions increases with respect to unstrained silicon because the effects of symmetry breaking due to strain increases the matrix elements and outweighs the decrease in available recombination pathways by the restriction of electron occupation to fewer valleys. 

\begin{center}
\textbf{Phonon Analysis for AMR}
\end{center}

\begin{figure}[ht]
\centering
\includegraphics[width=\linewidth,trim={0cm 0cm 0cm 0cm},clip]{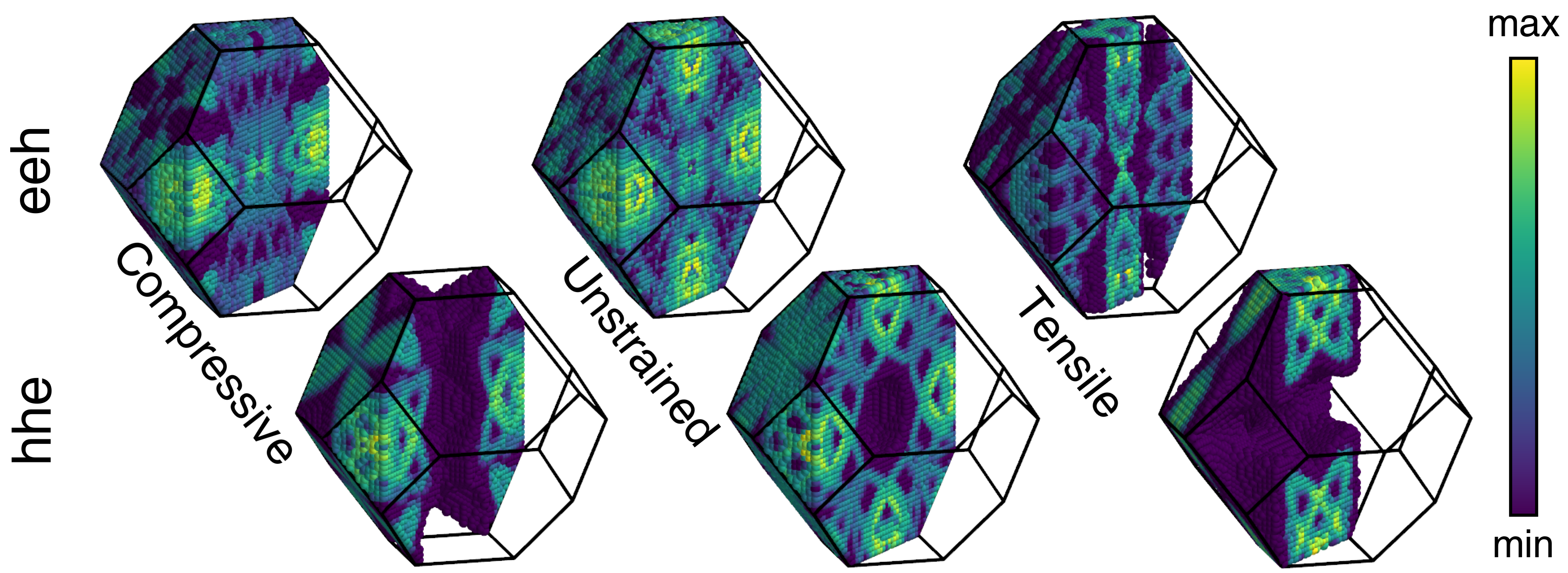}
\caption{The contributions by different phonon wave vectors throughout the first BZ for the \pa \(eeh\) and \(hhe\) AMR processes as a function of strain conditions. Brighter coloration indicates a stronger contribution to the \pa AMR process.}
\label{fig:strain_ph_bz}
\end{figure}

While we present the AMR distribution over phonon energies and wave-vector magnitudes in the main text, it is also  instructive to examine the three-dimensional distribution over phonon wave vectors in Fig. \ref{fig:strain_ph_bz}. The central (unstrained) figures are consistent with our previous study,\cite{Bushick2023} while the compressive and tensile plots show the anisotropy now present in the participatory phonons {\textendash} the compressive strain condition is now dominated by phonons with primarily in-plane wave vectors, while the tensile strain condition shows stronger selectivity for out-of-plane phonon wave vectors. These preferential selections are consistent with the significant increase in the potency of the \textit{g}-type process under strain. 

\begin{center}
\textbf{Excited Carrier Distributions}
\end{center}

\begin{figure}[ht]
\centering
\includegraphics[width=\linewidth,trim={0cm 0cm 0cm 0cm},clip]{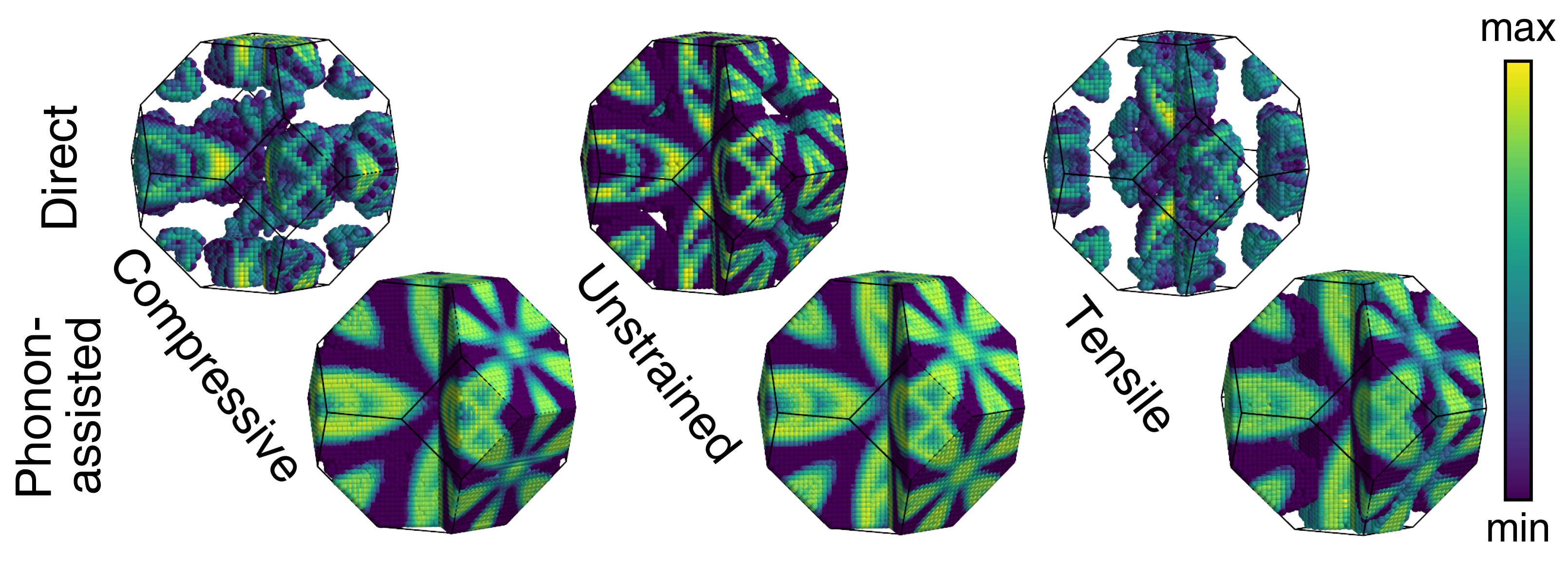}
\caption{The distribution of AMR-excited electrons throughout the first BZ for the direct and \pa \(eeh\) processes under compressive and tensile strain compared to the unstrained condition. A cutaway is made to show the internal structure. The anisotropy of the low-energy electrons paired with momentum conservation constraints gives rise to the anisotropy of the excited direct \am electrons, while the \pa \am electrons are able to scatter to states throughout the entire BZ.}
\label{fig:strain_k4_bz}
\end{figure}

Our methodology also allows us to examine the BZ distribution of the excited carriers. We find that the \pa distributions under all strain conditions are nearly equivalent for both the \(eeh\) and the \(hhe\) processes. This finding can be explained because strain affects the energy offsets at the VBM and CBM by \(\sim\)150 meV, but the excited carrier is promoted to a state \(\sim\)1 eV into the band. Given that the additional momentum provided by phonons allows carriers to effectively access the entire momentum space, we do not expect significant alterations to the excited carrier distributions. On the other hand, the direct \(eeh\) process is more strictly bounded by momentum conservation, and given the anisotropy in the starting electron distribution, we do find a consistent anisotropy in the excited electron distribution. We show both the direct and \pa excited electron distributions for the \(eeh\) process in Fig. \ref{fig:strain_k4_bz}, omitting the \(hhe\) figures since only the \pa process is relevant and the distribution of holes is consistent across the strain conditions. Comparing to the unstrained case, we demonstrate that, even though strain does limit the access of direct \(eeh\) AMR to some excited states due to lack of momentum conservation, the features of the total excited electron distribution (which is dominated by the \pa contribution) remain qualitatively similar to the unstrained case. 

\end{document}